%% file: freddice_3.tex
\pdfoutput=1
\RequirePackage{fix-cm}
\documentclass[reqno]{amsproc}
\usepackage{amssymb}
\usepackage{hyperref}
\usepackage{microtype}
\usepackage{lmodern}

\usepackage{tikz}
\usetikzlibrary{decorations.pathreplacing}

\usepackage{graphicx}

\usepackage[citation-order,nobysame]{amsrefs}
\usepackage{xyzbib}

\usepackage{mathtools}
\mathtoolsset{showonlyrefs}

\numberwithin{equation}{section}
\let\cite=\cites

\newtheorem{proposition}{Proposition}[section]
\newtheorem{conjecture}{Conjecture}[section]
\newtheorem{lemma}{Lemma}[section]

\DeclareMathOperator*{\tr}{\mathrm{tr}}
\DeclareMathOperator*{\res}{\mathrm{res}}
\DeclareMathOperator*{\diag}{\mathrm{diag}}
\DeclareMathOperator*{\Ai}{\mathrm{Ai}}

\let\leq=\leqslant

\renewcommand{\Re}{\mathrm{Re}\,{}}

\newcommand{\rme}{\mathrm{e}}
\newcommand{\rmi}{\mathrm{i}}
\newcommand{\rmd}{\mathrm{d}}

\newcommand{\wt}{\widetilde}

\newcommand{\yc}{y_\mathrm{c}}

\begin{document}

\vspace*{0in}

\title{Evaluation of integrals for the emptiness formation probability 
in the square-ice model}

\author{F. Colomo}
\address{INFN, Sezione di Firenze
Via G. Sansone 1, I-50019 Sesto Fiorentino (FI), Italy}
\email{colomo@fi.infn.it}

\author{A. G. Pronko}
\address{Steklov Mathematical Institute, 
Fontanka 27, 191023 Saint Petersburg, Russia}
\email{a.g.pronko@gmail.com}

\begin{abstract}  
We study the emptiness formation probability (EFP) in the six-vertex
model with domain wall boundary conditions. We present a conjecture
according to which at the ice point, i.e., when all the Boltzmann
weights are equal, the known multiple integral representation (MIR)
for the EFP can be given as a finite-size matrix determinant of
Fredholm type. Our conjecture is based on the explicit evaluation of
the MIR for particular values of geometric parameters and on two kinds
of identities for the boundary correlation function. The obtained
representation can be further written as the Fredholm determinant of
some linear integral operator.  We show that as the geometric
parameters of the EFP are tuned to the vicinity of the arctic curve
arising in the scaling limit, the conjectured determinant turns into
the GUE Tracy--Widom distribution.

\end{abstract}

\maketitle
\setcounter{tocdepth}{2}
\tableofcontents
\section{Introduction}

A useful tool in study of the six-vertex model with domain wall
boundary conditions is the `emptiness formation probability' (EFP), a
non-local correlation function describing the probability of obtaining
a region of the lattice with all vertices in the same state. It can be
seen as a cumulative distribution function with respect to the
geometric parameters describing the size of the `frozen' region. The
name `EFP' originated in the context of Heisenberg spin chains
(equivalent to the six-vertex model on a torus), where a similar
correlation function deserved a lot of attention, see, e.g.,
\cite{KBI-93,BK-02,BKNS-02,KLNSh-03,BKS-03,GKS-05,KKMST-09}.

In \cite{CP-07b}, we have derived various representations for the EFP
of the six-vertex model with domain wall boundary conditions, in terms
of determinants, orthogonal polynomials, and multiple contour
integrals. The multiple integral representation (MIR) has turned out
to be most useful, as it allowed us to derive the arctic curve of the
model \cite{CP-08,CP-09}.  The proposed approach found later a
simplified formulation with the tangent method \cite{CS-16}. As a
further development toward better understanding of the structures
arising in non-local correlation functions, an alternative derivation
of the known MIR based on certain antisymmetrization relation have
been provided \cite{CCP-19}. Recently, it was been shown that this
antisymmetrization relation can be used to obtain a new, alternative,
MIR for the EFP \cite{CGP-21}.  Furthermore, it appears that the
equivalence of the two MIRs implies for the boundary one-point
function entering these MIRs the existence of nontrivial relations,
which can be viewed as some sort of `sum rule' identities.

In the present paper, to elaborate further on these results, we focus
on a very interesting special case, known as the `ice point', where
all the  six-vertex model Boltzmann weights are equal to each other.
We find, that besides the sum rule type identities, there exist one
more type of relations for the one-point boundary correlation
function, that follows from the fact that at the ice point it can be
given in terms the Gauss hypergeometric function. Our main result is
an explicit, although conjectural expression for the EFP, as a
finite-size matrix determinant of the Fredholm type. Our conjecture is
based on the direct evaluation of the contour integrals in the first
MIR, and the above mentioned two types of identities. Next we
investigate the Fredholm determinant of the corresponding integral
operator. We show that, as the system size becomes large, and the
geometric parameters of the EFP are tuned to the vicinity of the
arctic curve, and appropriately scaled, the kernel of this integral
operator turns into the so-called Airy kernel.

It is well known that the Fredholm determinant of the Airy kernel
determines the celebrated Tracy--Widom distribution describing
fluctuations of the largest eigenvalue of the Gaussian unitary
ensemble (GUE) \cite{TW-94,M-04}. The Tracy--Widom distribution has
been observed to arise in a variety of models and it is commonly
believed to describe the crossover between weakly- and
strongly-coupled phases in various probabilistic problems, see,
e.g., \cite{D-06} and reference therein.  In particular, it has been
shown to arise in the discrete random matrix model with Hahn
polynomials type measure \cite{J-00} and in the closely related domino
tilings of the Aztec diamond \cite{J-02,J-05}.  The quantity studied
there can be viewed as the EFP of the six-vertex model with domain
wall boundary conditions and with Boltzmann weights restricted by the
free-fermion condition \cite{CP-13}. Appearance of the Tracy--Widom
distribution in the free-fermion six-vertex model context can also be
derived by means of a correspondence with non-intersecting
paths \cite{FS-06} and from the theory of Painlev\'e
equations \cite{KP-16}.

An important open problem actively addressed recently is how far this
law remains valid away from the free-fermionic case. Examples where it
has been shown to hold are the asymmetric simple exclusion
process \cite{TW-08a,TW-08b,TW-09} and the six-vertex model with
Boltzmann weights restricted by the stochastic
condition \cite{BCG-16}.  As for the general six-vertex model,
numerical simulations suggest a positive answer as
well \cite{LKV-23,PS-23}. A significant analytical achievement is the
proof that fluctuations for the maximum of the top path in
alternating-sign matrices, or, in other words, in the six-vertex model
at its ice point with domain wall boundary condition, are governed by
the Tracy--Widom distribution for the Gaussian orthogonal ensemble
(GOE) \cite{ACJ-23} (see Theorem 2.4 therein). As conjectured in that
paper, this would suggest the validity of the GUE Tracy--Widom
distribution for the EFP (Conjecture 2.5 therein).  The present paper
is strongly inspired by this observation.  Our main result here leads
exactly to the same statement, i.e., that the behaviour of the EFP of
the ice model with domain wall boundary conditions, in the vicinity of
the arctic curve, is described by the GUE Tracy--Widom distribution.

The paper is organized as follows. In the next section we collect all
necessary ingredients for calculations from the previous studies and
explain the origin of the two types of identities for the boundary
one-point function.  In section 3, we explain how we compute integrals
and use these identities to simplify the expressions, and provide the
main result that the EFP is given by a Fredholm-type determinant of a
finite-size matrix, see Conjecture~3.1. In section 4, we rewrite the
conjectured result in the form of the Fredholm determinant of a linear
integral operator, study it in the limit of large system size, and
show that when geometric parameters are tuned to the vicinity of the
arctic curve, and suitably rescaled, the kernel of the integral
operator turns into the Airy kernel, and thus the EFP into the GUE
Tracy--Widom distribution.

\section{MIRs for EFP}

In this section we collect all the necessary input information 
about the model: configurations of the model, 
the one-point boundary correlation function, two types 
of identities, and MIRs for the EFP.  

\subsection{The state sum}

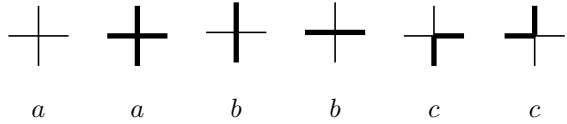
\begin{figure}
\centering
\input{fig-SixVertices}
\caption{The six vertices of the six-vertex model in terms of  
lines and their Boltzmann weights.}
\label{fieg-SixVertices}
\end{figure}

Configurations of the six-vertex model are usually depicted in terms
of arrows aligned along edges of a square lattice (or a four-valent
graph).  Allowed configurations are subject to the `ice rule': the
number of incoming and outgoing arrows at each lattice vertex is
equal.  Sometimes it is more convenient to use a description of states
in terms of occupation number variables, namely, an edge is considered
`empty' if it carries an up or right arrow, and `occupied' if it
carries a left or down arrow. Graphically, it corresponds to drawing a
solid lines on the lattice. The ice rule guaranties that the lines
flow from the top and right to the down and left. The six typical
vertices and their Boltzmann weights in the arrow-reversal symmetric
model are shown in Fig.~\ref{fieg-SixVertices}.

We consider the six-vertex model with homogeneous (vertex position
independent) Boltzmann weights. The three weight functions $a$, $b$,
and $c$ denoting the Boltzmann weights in the standard notation (see,
e.g. \cite{B-82}, for a detailed description) can be parameterized
(modulo overall normalization) by two parameters
\begin{equation}
\Delta=\frac{a^2+b^2-c^2}{2ab}, \qquad t=\frac{b}{a}.
\end{equation}  
The so-called ``ice point'' corresponds to all weights equal, $a=b=c$,
that is $\Delta=1/2$ and $t=1$. Our main results will be obtained for
this case, though some considerations appear to be valid for arbitrary
$\Delta$ and $t$.

Domain wall boundary conditions \cite{K-82,I-87,ICK-92} can be imposed
for the six-vertex model defined on a square domain of the square
lattice, consisting of $N$ horizontal and $N$ vertical lines (the
`$N\times N$ lattice'). With these conditions, each of the
four boundaries has all its edges carrying the same states and these
states are opposite to each other on opposite boundaries.

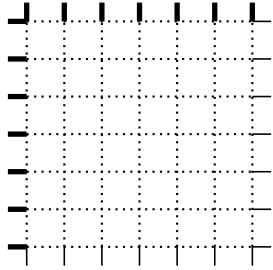
\begin{figure}
\centering
\input{fig-DWBCgrid}
\caption{Domain wall boundary conditions ($N=7$).
Configurations of the model are obtained by filling edges by thin or
thick lines.}
\label{fig-DWBCgrid}
\end{figure}

The partition function (the sum over states or `state sum') is
defined, as usual, as the sum over all configurations consistent with
the imposed boundary conditions, each configuration being assigned its
Boltzmann weight, given by the product of the weights of all
vertices. We will consider the partition function, normalized by the
factor $a^{N(N-1)}c^N$. We denote this quantity by $S_N$.  The chosen
normalization ensures that $S_N=S_N(t,\Delta)$ is a polynomial in $t$
and $\Delta$.  The first few values are:
\begin{align}
S_1&=1,
\\
S_2&=1+t^2,
\\
S_3&=\left(1+t^2\right)^3-2\Delta t^3,
\\
S_4&=\left(1+t^2\right)^6-8 \Delta  \left(1+2 t^2+2 t^4+t^6\right)t^3 
+4 \Delta^2\left(1+t^4\right)t^4.
\end{align}
At the ice point, $t=1$ and $\Delta=1/2$, $S_N$ is equal to the number
of $N\times N$ alternating-sign matrices, $S_1=1$, $S_2=2$, $S_3=7$,
$S_4=42$, etc. In general,
\begin{equation}
S_N(1,1/2)=A_N,\qquad A_N=\prod_{j=0}^{N-1}\frac{(3j-1)!}{(2N-j)!}.
\end{equation}
On the connection with the alternating-sign matrices,
see \cite{EKLP-92a,EKLP-92b,Ku-96,Ze-96,Br-99}.

\subsection{Boundary one-point function}

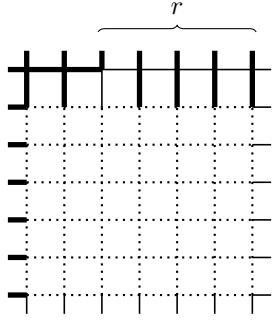
\begin{figure}
\centering
\input{fig-OnePointBCF}
\caption{Boundary one-point correlation function ($N=7$, $r=5$).}
\label{fig-OnePointBCF}
\end{figure}

A simple but important correlation function in the model is the
so-called boundary one-point function, originally introduced in
\cite{BPZ-02}. It is usually denoted by  $H_N^{(r)}$, and gives 
the probability that the $r$th vertical edge (e.g., from the right) in
the first horizontal row of the $N\times N$ lattice contains an
altered (e.g., empty vs occupied) state, see
Fig.~\ref{fig-OnePointBCF}. Note that because of the boundary
conditions there is exactly one such a state in this row (in general,
the $n$th row from the boundary contains $n$ altered states).

It is customary to deal with the related generating function of the
boundary one-point correlation function,
\begin{equation}\label{hNz-def}
h_N(z)=\sum_{r=1}^N H_N^{(r)} z^{r-1},\qquad h_N(1)=1.
\end{equation}
The function $h_N(z)$ admits an interpretation as the Izergin-Korepin
partition function \cite{I-87,ICK-92} with just one inhomogeneity
parameter \cite{S-06,CP-07b}.

To write down a general formula for $h_N(z)$, we introduce a sequence
of functions $\phi_j(t)$, $j=0,1,2,\ldots$, defining them in a
recursive way as follows \cite{PP-19}:
\begin{equation}
\phi_{j+1}(t)=t\partial_t (t-2\Delta+t^{-1}) \phi_{j}(t),\qquad \phi_0\equiv 1.
\end{equation}
Then, 
\begin{align}
h_N(z)
&
=(N-1)!\left((t-2\Delta +t^{-1})\frac{z}{z-1}\right)^{N-1}
\\ &\quad\times
\frac{\det\left[
\begin{cases}
\phi_{i+j-2}(t) & j=1,\ldots,N-1 \\
\phi_{i-1}(tz) & j=N 
\end{cases}
\right]_{i,j=1,\ldots,N} }{\det[\phi_{i+j-2}(t)]_{i,j=1,\ldots,N}}.
\end{align}
The first few functions $h_N(z)$ are:
\begin{align}
h_1(z)&=1,
\\
h_{2}(z)&=\frac{1+t^2 z}{S_2},
\\
h_3(z)&=
\frac{S_2+ 2\left(1+t^2-\Delta t\right)t^2 z+S_2 t^4 z^2}{S_3},
\\
h_4(z)&=\frac{1}{S_4}
\Big\{S_3+\left[3\left(1+t^2\right)^3t^2-2\Delta \left(3+6t^2+2t^4\right)t^3
+4\Delta^2t^4\right]z
\\ &\qquad
 +\left(3\left(1+t^2\right)^3t^2-2\Delta \left(2+6t^2+3t^4\right)t^3
 +4\Delta^2t^4\right)t^4z^2+ S_3 t^6 z^3\Big\}.
\end{align}
Here, $S_i$'s are the state sums given above. 

In some special cases the function $h_N(z)$ can be written explicitly
for generic $N$. For example, for $\Delta=0$, known as the
free-fermion point of the model, $h_N(z)=[(1+t^2 z)/(1+t^2)]^{N-1}$.

At the ice point, $t=1$ and $\Delta=1/2$, it
reads \cite{CP-08}
\begin{equation}\label{hNz-F1z}
h_N(z)={}_2F_1
\left(\genfrac{}{}{0pt}{}{-N+1,\, N}{2N} \bigg\vert 1-z\right).
\end{equation} 
or
\begin{equation}\label{hNz-Fz}
h_N(z) = \frac{(N)_{N-1}}{(2N)_{N-1}}
\, {}_2F_1
\left(\genfrac{}{}{0pt}{}{-N+1,\, N}{-2N+2} \bigg\vert z\right).
\end{equation}
It can also be written explicitly at the `dual ice point', $t=1$ and
$\Delta=-1/2$, though the expressions are rather bulky,
see \cite{CP-05a} for details.

The coefficients $H_N^{(r)}$ in \eqref{hNz-def} in the ice-point case
($t=1,\Delta=1/2$) can also be represented as
\begin{equation}
H_N^{(r)}= \frac{A_{N,r}}{A_N}.
\end{equation}
Here, $A_N$ is number of alternating-sign matrices of size $N\times
N$, and $A_{N,r}$ is their refined enumeration, i.e., the number of
such matrices with the sole `$1$' entry in the first row being at the
$r$th position \cite{Ze-96}.  In a more explicit form,
\begin{equation}
H_N^{(r)} = \binom{N+r-2}{N-1}\binom{2N-1-r}{N-1}\bigg/\binom{3N-2}{N-1},
\end{equation} 
that can be seen as another form of writing \eqref{hNz-F1z}
or \eqref{hNz-Fz}.

\subsection{Two types of identities}

To shorten the formulas a bit, we will skip the argument of functions
whenever it is $0$, e.g., writing $h_N$ for $h_N(0)$, $h_N'$ for
$h_N'(0)$, etc.

In the first type of identities the derivatives of $h_N(z)$ at the
point $z=1$ are expressed in terms of the derivatives of $h_N(z)$ at
the point $z=0$. The first three relations read:
\begin{equation}\label{SumRules}
\begin{split}
h_{N-1}'(1)
&=\frac{1}{1-2\Delta t+t^2}\left\{\frac{h_N'}{h_N}-t^2\right\},
\\ 
h_{N-2}''(1)
&=\frac{1}{(1-2\Delta t+t^2)^2}\bigg\{-\frac{h_N''}{h_N}
+2\frac{h_{N-1}'h_N'}{h_{N-1}h_N}
-2\left(1-2\Delta t +2t^2\right)\frac{h_{N-1}'}{h_{N-1}}
\\ &\qquad
+2\frac{h_N'}{h_N}
-2t^2+2t^4
\bigg\},
\\ 
h_{N-3}'''(1)
&=
\frac{1}{(1-2\Delta t+t^2)^2}\bigg\{
\frac{h_N'''}{h_N}-3\frac{h_{N-2}' h_N''}{h_{N-2} h_N}
-3 \frac{h_{N-1}'' h_N'}{h_{N-1} h_N}
\\ &\qquad
+3 \left(2+3 t^2-4 t \Delta \right) \frac{h_{N-1}''}{h_{N-1}}
-6 \frac{h_N''}{h_N}
+6\frac{h_{N-2}' h_{N-1}' h_{N}'}{h_{N-2} h_{N-1} h_{N}}
\\ &\qquad
-6 \left(2+3 t^2-4 t \Delta\right)  \frac{ h_{N-2}'h_{N-1}'}{h_{N-2}h_{N-1} }
+6 \frac{h_{N-2}'h_N' }{ h_{N-2}h_N}
+6 \frac{h_{N-1}'h_N'}{h_{N-1}h_N }
\\ &\qquad
+6 \left(1+2 t^2+3 t^4-4 t \Delta -6 t^3 \Delta +4 t^2 \Delta^2\right)
\frac{h_{N-2}'}{h_{N-2}}
\\ &\qquad
-6 \left(2+3 t^2-4 t \Delta \right)\frac{ h_{N-1}'}{h_{N-1}}
+6 \frac{h_N'}{h_N}+18 t^4-6 t^6-12 t^3 \Delta
\bigg\}.
\end{split}
\end{equation}
In view of \eqref{hNz-def}, these identities express sums over the set
of functions $H_N^{(r)}$, $r=1,\ldots,N$, in terms of first few of
them, i.e., they are sum rule identities. As explained in the next
section, these identities follows from the existence of two different
MIRs for the EFP.

Another sort of identities involves only higher derivatives at
$z=0$. It has to be stressed, however, that we are able at the moment
to establish them only for the special case of the ice point, $t=1$
and $\Delta=1/2$.  In this case one can use the explicit
expression \eqref{hNz-Fz}.

For example, from \eqref{hNz-Fz} one can easily find for the first
derivative of $h_N(z;1,1/2)$ at $z=0$:
\begin{equation}
\frac{h_N'}{h_N} =\frac{N}{2}.
\end{equation}
Hence, we have 
\begin{equation}\label{h'h}
\frac{h_N'}{h_N}-\frac{h_{N-1}'}{h_{N-1}}-\frac{1}{2}=0.
\end{equation}
Clearly, relation \eqref{h'h} shows that the identities in \eqref{SumRules} 
at the ice point can be 
somewhat simplified, e.g., by expressing 
$h_{N-1}'/h_{N-1}$, $h_{N-2}'/h_{N-2}$, etc, in terms of $h_{N}'/h_{N}$.

For the second derivative we get 
\begin{equation}
\frac{h_N''}{h_N} =\frac{(N-2)N(N+1)}{2(2N-3)}.
\end{equation}
Taking into account that
\begin{equation}
\frac{h_{N-1}}{h_{N}}=\frac{3(3N-2)(3N-4)}{4(2N-1)(2N-3)}
\end{equation}
one can find the following identity:
\begin{equation}\label{h''h}
\frac{h_N''}{h_N}-\frac{h_{N-1}''}{h_{N-1}}-\frac{h_N'}{h_N}
-2\frac{h_{N-2}}{h_{N-1}}+\frac{7}{2}=0.  
\end{equation}  
Note that the coefficients are all independent of $N$. 

For the third derivative we have
\begin{equation}
\frac{h_N'''}{h_N}=\frac{(N-3)N(N+1)(N+2)}{4(2N-3)}.
\end{equation}
In the similar manner, we get
\begin{equation}
\frac{h_N'''}{h_N}-\frac{h_{N-1}'''}{h_{N-1}}
-\frac{3}{2}\left(\frac{h_N'}{h_N}\right)^2
-\frac{21}{2}\left(\frac{h_{N-2}}{h_{N-1}}-\frac{7}{4}\right)=0.  
\end{equation}
Clearly, this game can be continued. 

Note that the first type of identities make it possible to express
$h_{N-s}^{(s)}(1)$ in terms of the quantities $h_{N-k}^{(j)}$, with
$j=0,\dots,s$ and $k=0,\dots, s-1$, such that $0\leq j+k \leq s$. In
turn, the second type of identities make it possible to express these
quantities in terms of
$h_{N-s+1},\ldots,h_{N},h_{N}',\ldots,h_{N}^{(s)}$, thus reducing the
number of formally independent objects from $s(s+3)/2$ to $2s$.

\subsection{EFP}

The EFP, denoted as $F_N^{(r,s)}$, can be defined as the probability
of obtaining a domain of size $s\times (N-r)$ attached to a corner of
the $N\times N$ lattice, with all its horizontal and vertical edges
having the same states as those at the boundary. In other words, all
vertices belonging to this $s\times (N-r)$ domain are all $a$- or
$b$-weight vertices; this domain can be regarded as an `empty'
region.  The vertices belonging to this domain can be removed, giving
rise to so-called L-shaped domain. The partition function of the model
on this domain is exactly the EFP, modulo the factor $a^{s(N-r)}$ or
$b^{s(N-r)}$ times the partition function of the model on the original
$N\times N$ lattice.

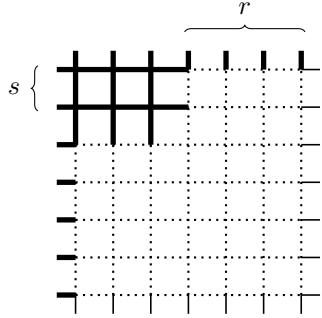
\begin{figure}
\centering
\input{fig-EFP}
\caption{Emptiness formation probability ($N=7$, $r=4$, $s=2$).}
\label{fig-EFP}
\end{figure}

A simple property of the EFP is that it vanishes identically whenever
$s>r$. This corresponds to a situation where the frozen rectangular
domain entering the definition of the EFP extends beyond the diagonal
of the $N\times N$ lattice. In terms of the model in the L-shaped
domain, it is clear that whenever $s>r$, the number of allowed
configurations vanishes.

Here, for definiteness we stick at the conventions used in
Refs.~\cite{CP-07b,CP-09,CGP-21}, and define EFP such that the frozen
domain is located at the top left corner of the $N\times N$ lattice,
with all vertices of this domain to be $a$-weight vertices, see
Fig.~\ref{fig-EFP}. In \cite{CP-07b}, the following MIR have been
derived:
\begin{multline}\label{efpMIR1}
F_N^{(r,s)} = (-1)^s
\oint_{C_0}^{} \cdots \oint_{C_0}^{}
\prod_{j=1}^{s}\frac{[(t^2-2\Delta t)z_j+1]^{s-j}}{z_j^r(z_j-1)^{s-j+1}}\,
\\ \times
\prod_{1\leq j<k \leq s}^{} \frac{z_j-z_k}{t^2z_jz_k-2\Delta t z_j+1}\;
h_{N,s}(z_1,\dots,z_s)
\,\frac{\rmd^s z}{(2\pi \rmi)^s}.
\end{multline}
Here, $C_0$ denotes a small simple anticlockwise oriented contour
around the point $z=0$.  The functions $h_{N,s}(z_1,\dots,z_s)$,
$s=1,\ldots,N$, are symmetric polynomials of the degree $N-1$ in each
of their variables and defined in terms of $s$ single-variable
functions $h_{N-s+1}(z),\ldots,h_N(z)$ by the formula
\begin{equation}\label{hNs}
h_{N,s}(z_1,\dots,z_s) =
\frac{\det
\left[(z_j-1)^{s-i} z_j^{i-1}h_{N-i+1}(z_j)\right]_{i,j=1,\ldots,s}}
{\prod_{1\leq i<j \leq s}^{} (z_i-z_j)}.
\end{equation}
If, say, $z_s=1$, then
\begin{equation}
h_{N,s}(z_1,\dots,z_{s-1},1)=h_{N,s-1}(z_1,\dots,z_{s-1}).
\end{equation}
If $z_s=0$, then
\begin{equation}
h_{N,s}(z_1,\dots,z_{s-1},0)=h_N(0)h_{N-1,s-1}(z_1,\dots,z_{s-1})
\end{equation}
i.e., the degree in all the remaining variables decreases by one. 

In \cite{CGP-21},  it was proven that for generic values of 
$t$ and $\Delta$, together with MIR 
\eqref{efpMIR1}, EFP admits also the following MIR: 
\begin{multline}\label{efpMIR2}
F_N^{(r,s)}=
\frac{h_N(0)\cdots h_{N-s+1}(0)t^{(r-s)(r-s-1)}}{(r-s)! h_{1}(0)\cdots h_{r}(0)}
\oint_{C_1}\cdots\oint_{C_1} \prod_{j=1}^{r-s} \frac{1}{z_j-1}
\\ \times
\prod_{\substack{j,k=1\\j\ne k}}^{r-s}
\frac{z_j-z_k}{t^2z_jz_k-2\Delta t z_j+1}
\,
h_{N-s,r-s}\left(z_1,\dots,z_{r-s}\right)
\\ \times
h_{r,r-s}\left(\frac{t^2z_1-2\Delta t +1}{t^2(z_1-1)},\dots,
\frac{t^2z_{r-s}-2\Delta t +1}{t^2(z_{r-s}-1)}\right)
\frac{\rmd^{r-s} z}{(2\pi\rmi)^{r-s}},
\end{multline}
where $C_1$ is a (small) contour encircling the point $z=1$.  In
comparison with \eqref{efpMIR1}, the number of integrations
in \eqref{efpMIR2} is $r-s$, that is the lattice distance of the point
$(r,s)$ from the antidiagonal, rather than $s$, the lattice distance
from the top boundary.

As far as MIRs \eqref{efpMIR1} and \eqref{efpMIR2} describe the same
quantity and both are valid for all allowed values of the geometric
parameters, they imply some relations for the function $h_N(z)$. These
relations are exactly the first type identities, \eqref{SumRules}.
Indeed, setting $r=s+1$, one gets for the second MIR just single
integration around the pole at the point $z=1$, while the first MIR
contains $s$ integrations around the poles at the points
$z_1,\ldots,z_s=0$. Considering the cases $s=1,2,3,\ldots$ one arrives
at the identities listed in \eqref{SumRules}.

\section{Evaluation of the integrals}

In this section we address the problem of evaluation of the MIR 
\eqref{efpMIR1} for $s=1,2,3,4$ and assuming that $r=N-s$, 
in the case of the ice point ($t=1, \Delta=1/2$). In principle, this
calculation can be done for generic values of the Boltzmann weights,
but at the present moment we able to formulate a conjecture only for
this special case.

\subsection{Deformation of the contours}

We start with making a change of the variables 
$z_j\mapsto z_j^{-1}$, $j=1,\ldots,s$ 
in \eqref{efpMIR1}, that gives 
\begin{equation}\label{efpMIRinfty}
F_N^{(r,s)} =
\oint_{C_\infty}^{} \cdots \oint_{C_\infty}^{}
J_N^{(r,s)}(z_1,\ldots,z_s)
\,\rmd^s z,
\end{equation}
where $C_\infty$ denotes a very large contour, anticlockwise oriented,
and
\begin{multline}\label{JNrs}
J_N^{(r,s)}(z_1,\ldots,z_s)
=(2\pi \rmi)^{-s}\prod_{j=1}^{s}
\frac{[t^2-2\Delta t+z_j]^{s-j}}{z_j^{N-r}(z_j-1)^{s-j+1}}\,
\prod_{1\leq j<k \leq s}^{} \frac{z_j-z_k}{t^2-2\Delta t z_k+z_jz_k}\;
\\ \times
\tilde h_{N,s}(z_1,\dots,z_s).
\end{multline}
Here, 
\begin{equation}\label{htNs}
\tilde h_{N,s}(z_1,\dots,z_s) \equiv (z_1\cdots z_s)^{N-1} h_{N,s}(z_1^{-1},
\dots,z_s^{-1}). 
\end{equation}
These functions can also be written as determinants, similarly
to \eqref{hNs},
\begin{equation}
\tilde h_{N,s}(z_1,\dots,z_s) =\frac{\det
\left[(z_j-1)^{s-i} z_j^{i-1}\tilde h_{N-i+1}(z_j)\right]_{i,j=1,\ldots,s}}
{\prod_{1\leq i<j \leq s}^{} (z_i-z_j)}.
\end{equation}
The function $\tilde h_N(z)$ is defined by
\begin{equation}
\tilde h_N(z)\equiv z^{N-1} h_N(z^{-1})
\end{equation}
and it can also be related to $h_N(z)$ by mapping $t\mapsto t^{-1}$,
\begin{equation}
\tilde h_N(z;t)=h_N(z;t^{-1}).
\end{equation}
Hence, in the symmetric case $t=1$ the tildes over the functions can
be lifted.

We continue transforming the expression by deforming the integration
contours, shrinking them down to encircle the sole poles at $z=0$ and
$z=1$. An important property of the integrand \eqref{JNrs} is that in
the deformation of the contours all terms deriving from the evaluation
of the residues at mutual poles (those due to the double product in
the denominator) vanish \cite{CGP-21}. As a result, we have for the
EFP
\begin{equation}\label{MIR1c1c0}
F_N^{(r,s)} = 
\oint_{C_1\cup C_0}^{} \cdots \oint_{C_1\cup C_0}^{}
J_N^{(r,s)}(z_1,\ldots,z_s)\,\rmd^s z.
\end{equation}
This formula implies that the EFP is essentially the sum of $s+1$
terms:
\begin{equation}
F_N^{(r,s)}=\sum_{k=0}^{s} I_k,
\end{equation}
where $I_k$ is the sum of integrals with $k$ variables integrated over
$C_0$ and with the remaining $s-k$ variables integrated over
$C_1$. The following two simple lemmas show that $I_0=1$ and establish
the value of $I_s$ in the case of our interest.

\begin{lemma}
For arbitrary values of all parameters, a cumulative residue at $z=1$ 
in \eqref{JNrs} is equal to one,
\begin{equation}
\res_{z_1=1}\ldots\res_{z_s=1 }J_N^{(r,s)}(z_1,\ldots,z_s)=1.    
\end{equation}  
\end{lemma}
\begin{proof}
To prove this statement, evaluate the residues in the shown order.  At
each step the pole in the corresponding variable at the point $z=1$ is
simple, and the result follows due to $h_{N,s}(1,\ldots,1)=1$.
\end{proof}

\begin{lemma}
At the ice point, $\Delta=1/2,t=1$, and for $N-r=s$ the cumulative
residue in \eqref{JNrs} at the point $z=0$ equals:
\begin{equation}
\res_{z_1=0}\ldots\res_{z_s=0}J_N^{(N-s,s)}(z_1,
\ldots,z_s)=(-1)^s h_N\cdots h_{N-s+1},
\end{equation} 
where $h_N\equiv h_N(0)$, etc.
\end{lemma}
\begin{proof}
The proof is similar to that of the previous lemma.  
\end{proof}

\subsection{Determinant structures}

Our strategy in computing the EFP is based on the observation (see
also \cite{TW-08b,STW-22}) that there exists an $s\times s$ matrix $A$
such that
\begin{equation}\label{detIA}
\sum_{k=0}^s \lambda^k I_k = {\det}_s (I- \lambda A),
\end{equation}
where $I$ is the $s\times s$ identity matrix and $\lambda$ is a formal
parameter.  In the examples below matrix $A$ is such that by
eliminating its last row and column, the reduction $s\mapsto s-1$,
$N\mapsto N-1$ is made. As we shall see, this reduction is a key
property of the expressions for $I_k$'s computed
from \eqref{MIR1c1c0}. We have also found that the matrix $A$ can be
given explicitly in a factorized form
\begin{equation}\label{adlu}
A=D\,L\,U,
\end{equation}
where $D$, $L$, and $U$ is a 
diagonal, lower-triangle, and upper-triangle matrix, respectively.

Representation \eqref{detIA} is not unique and one can also to try to
find a similar formula
\begin{equation}
\sum_{k=0}^s \lambda^k I_k = {\det}_s (I- \lambda V),
\end{equation}
where the matrix $V$ has entries independent of $s$. We find such a
matrix and it is in fact related to $A$. Matrix $V$ can be given as a
product of lower-triangle, diagonal, and upper-triangle matrices:
\begin{equation}\label{vldu}
V =\wt L\, \wt D\, \wt U.
\end{equation}
They appear to be related to $L$, $D$, and $U$ by 
\begin{equation}\label{tildeT-even}
\wt L = \varOmega L^\mathsf{T} \varOmega,\qquad  
\wt D = \varOmega D \varOmega,\qquad
\wt U = \varOmega U^\mathsf{T} \varOmega, 
\end{equation}
for $s$ even, and 
\begin{equation}\label{tildeT-odd}
\wt L = \varOmega U \varOmega,\qquad  
\wt D = \varOmega D \varOmega,\qquad
\wt U = \varOmega L \varOmega, 
\end{equation}
for $s$ odd. Here $\mathsf{T}$ denotes matrix transposition, and
$\varOmega$ is the $s\times s$ `unit anti-diagonal' matrix,
\begin{equation}
\Omega_{ij}=\delta_{i+j,s+1},\qquad 
\varOmega^\mathsf{T}=\varOmega,\qquad 
\varOmega^2=I.
\end{equation} 

The sum-rule relations \eqref{SumRules} are essentially used in
obtaining expressions for quantities $I_k$ in terms of the values of
derivatives of the function $h_N(z)$ at $z=0$. To find the matrix $A$,
starting from the $s=3$ case, we need to use also relations between
these values. These are the second type of identities; by reducing the
number of formally independent quantities from $s(s+1)/2$ to $2s-1$,
they play a crucial role in the construction of our main conjecture
about entries of the matrices $A$ and $V$.

To proceed with particular cases, we need some more notation. 
To shorten formulas below, we denote
\begin{equation}
b_i\equiv h_{N-i}, \qquad i=0,1,2,\ldots,s-1, 
\end{equation} 
and (we limit ourselves below in writing explicitly terms with three
derivatives)
\begin{equation}
\kappa_i'=\frac{h_{N-i}'}{h_{N-i}},\quad \kappa_i''=\frac{h_{N-i}''}{h_{N-i}},
\quad \kappa_i'''=\frac{h_{N-i}'''}{h_{N-i}}.
\end{equation}  
We recall that $h_N\equiv h_N(0)$, $h_N'\equiv h_N'(0)$,  etc.   

\subsection{Particular cases of $s=1,\ldots,4$}

We now turn to evaluation of the integrals in the case of
$s=1,\ldots,4$.  These expressions appear to be very instructive in
our conjecture of the result for a generic $s$, considered next.

\subsubsection{Case $s=1$}

We start with the simplest but useful case of $s=1$. Evaluating the
integral for $r=N-1$ and using that $\tilde h_N(1)=1$, we get for the
EFP (for generic $\Delta$ and $t$) the expression
\begin{equation}
F_N^{(N-1,1)}=1-\tilde h_N.
\end{equation}
At $t=1$ we have $\tilde h_N =h_N\equiv b_0$ and so 
\begin{equation}\label{Is=1}
I_0=1,\qquad I_1=-b_0.
\end{equation}
Hence,
\begin{equation}\label{AVb0}
A=V=b_0.
\end{equation}

\subsubsection{Case $s=2$} 

This is very instructive example and so we will be able to explain
main ideas of further calculations.  Evaluation of the integrals
in \eqref{MIR1c1c0} in this case, using that $\tilde h_N(1)=1$, yields
\begin{align}
I_0&=1,
\\
I_1&=-\frac{t^4-2\Delta t^3+4\Delta^2 t^2-1}{t^4}\,\tilde h_N
-\frac{2\Delta}{t}\, \tilde h_{N}'
-\frac{t^2-2\Delta t+1}{t^2}\,\tilde h_{N-1}
\\ &\qquad
-\left(\frac{2\Delta t-1}{t^2}\,  
\tilde h_{N}
+\tilde h_{N}'\right)\frac{t^2-2\Delta t+1}{t^2}\, \tilde h_{N-1}'(1),
\\
I_2&=\frac{t^2-4\Delta t+4\Delta^2+1}{t^2}
\tilde h_N \tilde h_{N-1}.
\end{align}
Now, setting $t=1$ and $\Delta=1/2$ (note that we must keep generic
$t$ or $\Delta$, or both, before evaluation of integrals, in order to
avoid erroneous contributions from the term $t^2-2\Delta t z_2+z_1z_2$
when computing $I_1$), and using the first identity
from \eqref{SumRules}, we get
\begin{equation}\label{Is=2}
I_0=1,\qquad I_1=-b_1-b_0(\kappa_0')^2,\qquad I_2=b_0b_1.
\end{equation}
Note that if we put here $b_0=0$, then we get the $s=1$
result \eqref{Is=1} in which $b_0$ is replaced by $b_1$, or $N\mapsto
N-1$.  This means that if an $2\times 2$ matrix $A$ exists such
that \eqref{detIA} holds, then it would be desirable that its top-left
entry is $b_1$.  In turn, this choice fixes the bottom-right entry
from the relation $\tr A=-I_1$ to be $b_0(\kappa_0')^2$.  The
off-diagonal entries must satisfy $\det A =I_2$, that modulo diagonal
similarity transformation (and up to matrix transposition) leads us to
\begin{equation}
A=
\begin{pmatrix}
b_1 & b_1(\kappa_0'-1)
\\
b_0(\kappa_0'+1) & b_0(\kappa_0')^2   
\end{pmatrix}. 
\end{equation}
Clearly, this matrix possesses the DLU-factorization \eqref{adlu},
where
\begin{equation}
D=
\begin{pmatrix}
b_1 & 0
\\
0 & b_0
\end{pmatrix}, 
\quad
L=
\begin{pmatrix}
1 & 0
\\
\kappa_0'+1& 1 
\end{pmatrix}, 
\quad
U=
\begin{pmatrix}
1 & \kappa_0'-1
\\
0 & 1   
\end{pmatrix}. 
\end{equation}

Another way to write the result of integration in a determinant form
is with a matrix $V$ such that $s$ determines only its size, but not
entries.  For such a matrix then, first from \eqref{AVb0} follows
$V_{11}=b_0$, next from $\tr V=-I_1$ follows
$V_{22}=b_0\left((\kappa_0')^2-1\right)+b_1$, and finally from $\det
V=I_2$ follows $V_{12}V_{21}=b_0^2\left((\kappa_0')^2-1\right)$.
Modulo diagonal transformation and matrix transposition,
\begin{equation}
V=
\begin{pmatrix}
b_0 & b_0(\kappa_0'-1)
\\
b_0(\kappa_0'+1) & b_0\left((\kappa_0')^2-1\right)+b_1   
\end{pmatrix}.
\end{equation}
This matrix admits LDU-factorization \eqref{vldu}, where
\begin{equation}
\wt L=
\begin{pmatrix}
1 & 0
\\
\kappa_0'+1& 1 
\end{pmatrix},
\quad 
\wt D=
\begin{pmatrix}
b_0 & 0
\\
0 & b_1
\end{pmatrix},
\quad 
\wt U=
\begin{pmatrix}
1 & \kappa_0'-1
\\
0 & 1   
\end{pmatrix}. 
\end{equation}
These matrices are related to $L$, $D$, and $U$
by \eqref{tildeT-even}.

\subsubsection{Case $s=3$}
In this case, after evaluating the integrals in MIR \eqref{MIR1c1c0},
using $\tilde h_N(1)=1$, applying the first two identities from
\eqref{SumRules}, and putting $t=1$, $\Delta=1/2$, we get 
\begin{equation}
\begin{split}\label{Is=3}
I_0&=1,
\\
I_1 & =-b_2-b_1 \left(\kappa_1'\right)^2-b_0
\left[\left(\frac{\kappa_0''}{2}-\kappa_0'\right)^2+2\kappa_0'-1\right],
\\
I_2 & = b_1b_2+b_0b_2
\left(\kappa_0'\right)^2+b_0b_1\left[1-\kappa_0'(1+\kappa_1')
+\frac{\kappa_0''}{2}\right]^2,
\\
I_3 & = -b_0b_1b_2.
\end{split}
\end{equation}
If we put $b_0=0$ in \eqref{Is=3} then we get 
\eqref{Is=2} in which $b_0, b_1,\kappa_0'\mapsto b_1, b_2,\kappa_1'$,
respectively, that is, $N\mapsto N-1$.

Thus, we can construct $A$ by choosing its top-left $2\times 2$ block
as the matrix $A$ from the case $s=2$, with $b_0, b_1,\kappa_0'\mapsto
b_1, b_2,\kappa_1'$, that also fixes $A_{33}$ entry from $\tr A=-I_1$.
We write $A=DLU$, where $D=\diag(b_2,b_1,b_0)$, and for the matrices
$L$ and $U$ we take
\begin{equation}
L=
\begin{pmatrix}
1 & 0 & 0\\
\kappa_1'+1 & 1 & 0
\\ 
\frac{1}{2}\kappa_0''-2\kappa'_0+1 & \kappa_0'-1& 1
\end{pmatrix}, 
\qquad
U=
\begin{pmatrix}
1 & \kappa_1'-1 & \frac{1}{2}\kappa_0''-1\\
0 & 1 & \kappa_0'+1
\\ 
0 & 0& 1
\end{pmatrix}.
\end{equation}
Clearly, it can easily seen that this choice reproduces values $I_0$,
$I_1$ and $I_3$, but not $I_2$, with the difference of some
complicated factor times $\kappa_1'-\kappa_0'+\frac{1}{2}$. But it
vanishes due to identity \eqref{h'h}!

The similar construction in the case of the matrix $V$, whose entries 
are required to be independent of $s$, leads us to \eqref{vldu},
where $\wt D=\diag(b_0,b_1,b_2)$ and   
\begin{equation}
\wt L=
\begin{pmatrix}
1 & 0 & 0\\
\kappa_0'+1 & 1 & 0
\\ 
\frac{1}{2}\kappa_0''-1 & \kappa_1'-1& 1
\end{pmatrix}, 
\qquad
\wt U=
\begin{pmatrix}
1 & \kappa_0'-1 & \frac{1}{2}\kappa_0''-2\kappa_0'+1\\
0 & 1 & \kappa_1'+1
\\ 
0 & 0 & 1
\end{pmatrix}. 
\end{equation}
These matrices are related to $D$, $L$, $U$ by \eqref{tildeT-odd}.
 
\subsubsection{Case $s=4$} In this case, we obtain
\begin{align}
I_0&=1,
\\ 
I_1&=
-b_3-b_2 \left(\kappa_2'\right)^2
+b_1 \left[1-2 \kappa_1'-\left(\kappa_1'-\frac{\kappa_1''}{2}\right)^2 \right]
\\ &\quad
+b_0
\left[2-6 \kappa_0'+3 \left(\kappa_0'\right)^2+\left(1-\kappa_0'\right) \kappa_0''
-\left(\kappa_0'-\kappa_0''+\frac{\kappa_0'''}{6}\right)^2\right],
\\
I_2&=
b_2 b_3+b_1 b_3 
\left(\kappa_1'\right)^2
+b_0 b_3 \left[-1+2 \kappa_0'+\left(\kappa_0'-\frac{\kappa_0''}{2}\right)^2\right]
\\ &\quad
+b_1 b_2 \left[1-\kappa _1' \left(1+\kappa _2'\right)+\frac{\kappa _1''}{2}\right]^2
\\ &\quad 
+b_0 b_2
\Bigg\{-\left(5+\kappa_2'\right)\left(1+\kappa_2'\right)+2\kappa_0' 
\left(1+\kappa_2'\right)^2
\\ &\qquad\qquad
+\left[1+\left(2+\kappa
   _2'\right) \left(-\kappa _0'+\frac{\kappa _0''}{2}\right)-\frac{\kappa_0'''}{6}\right]^2
   \Bigg\}
\\ &\quad
+b_0 b_1 \Bigg\{4-2 \kappa_0'+\frac{\kappa_0'''}{3}
+\left(1-\kappa_0'+\frac{\kappa_0''}{2}\right)^2
\\ &\qquad\qquad
+2 \kappa _1' \left[2+\left(2-\kappa_0'+\frac{\kappa_0''}{2}\right) 
\left(1-\kappa_0'+\frac{\kappa_0''}{2}-\frac{\kappa_0'''}{6}\right)\right]
\\ &\qquad\qquad
+\kappa_1''
   \left[\left(\kappa_0'-\frac{\kappa_0''}{2}\right)^2
   -3 \kappa_0'+\frac{3 \kappa_0''}{2}-\frac{\kappa_0'''}{6}\right]
\\ &\qquad\qquad
+\left(\kappa_1'\right)^2 \left(1-\kappa _0'+\frac{\kappa_0''}{2}
-\frac{\kappa_0'''}{6}\right)^2
\\ &\qquad\qquad
+\left(\frac{\kappa_1''}{2} \right)^2 
\left[-1+2 \kappa _0'+\left(\kappa _0'-\frac{\kappa
   _0''}{2}\right)^2\right]
\\ &\qquad\qquad
-\kappa_1' \kappa_1'' \left[2+\left(\kappa _0'-\frac{\kappa _0''}{2}\right) \left(1-\kappa
   _0'+\frac{\kappa _0''}{2}-\frac{\kappa_0'''}{6}\right)\right]\Bigg\},
\\
I_3 &=
-b_1 b_2 b_3-b_0 b_2 b_3 
\left(\kappa_0'\right)^2
-b_0 b_1 b_3 \left[1-\kappa_0' \left(1+\kappa_1'\right)
+\frac{\kappa_0''}{2}\right]^2
\\ &\quad
-b_0 b_1 b_2 
\left\{1+\kappa_0''+\kappa_2' \left(1+\frac{\kappa_0''}{2}\right)
-\kappa_0'\left[\left(1+\kappa_1'\right) \left(1+\kappa_2'\right)-\frac{\kappa_1''}{2}
\right]-\frac{\kappa_0'''}{6}\right\}^2,
\\ 
I_4&= b_0b_1b_2b_3.
\end{align}
Repeating the procedure from the $s=2$ and $s=3$ cases, we write 
matrix $A$ in the form \eqref{adlu}, with 
\allowdisplaybreaks
\begin{align}
D&=\diag(b_3,b_2,b_1,b_0),
\\
L&
=\begin{pmatrix}
1 & 0 & 0 & 0\\
 \kappa_2'+1& 1 & 0 & 0
\\ 
 \frac{1}{2}\kappa_1''-2\kappa_1'+1 
& \kappa_1'-1& 1 & 0 
\\
\frac{1}{6}\kappa_0'''-\frac{1}{2}\kappa_0''-\kappa_0'+1
&\frac{1}{2}\kappa_0''-1 & \kappa_0'+1 & 1
\end{pmatrix}, 
\\
U&
=\begin{pmatrix}
1 &  \kappa_2'-1& \frac{1}{2}\kappa_1''-1&
\frac{1}{6}\kappa_0'''-\frac{3}{2}\kappa_0'' +3\kappa_0'-1
\\
0 & 1 & \kappa_1'+1 & \frac{1}{2}\kappa_0''-2\kappa_0'+ 1
\\ 
0 & 0& 1 & \kappa_0'-1 
\\ 
0 & 0 & 0 & 1
\end{pmatrix}. 
\end{align}
\allowdisplaybreaks[0]%
This choice indeed works, but one have to use the identity \eqref{h'h}
to express now $\kappa_1'$ and $\kappa_2'$, for the latter twice, to
get them both expressed in terms of $\kappa_0'$. Furthermore, the
identity \eqref{h''h} is also needed, to express $\kappa_1''$ in terms
of $\kappa_0''$, $b_2$ and $b_1$, which are already involved.

As for the matrix $V$ we get \eqref{vldu}, where
\begin{align}
\wt L&
=\begin{pmatrix}
1 & 0 & 0 & 0\\
\kappa_0'+1 & 1 & 0 & 0
\\ 
\frac{1}{2}\kappa_0''-1 & \kappa_1'-1& 1 & 0 
\\
\frac{1}{6}\kappa_0'''-\frac{1}{2}\kappa_0''-\kappa_0'+1
& \frac{1}{2}\kappa_1''-2\kappa_1'+1 & \kappa_2'+1 & 1
\end{pmatrix}, 
\\
\wt D&=\diag(b_0,b_1,b_2,b_3),
\\
\wt U&=\begin{pmatrix}
1 & \kappa_0'-1 & \frac{1}{2}\kappa_0''-2\kappa_0'+ 1&
\frac{1}{6}\kappa_0'''-\frac{3}{2}\kappa_0'' +3\kappa_0'-1
\\
0 & 1 & \kappa_1'+1 & \frac{1}{2}\kappa_1''-1
\\ 
0 & 0& 1 & \kappa_2'-1
\\ 
0 & 0 & 0 & 1
\end{pmatrix}. 
\end{align}
These matrices are related to $L$, $D$, and $U$ by \eqref{tildeT-even}.

\subsection{Case of generic $s$}

To formulate a conjecture about entries of the matrices $L$ and $U$ 
(while this is not difficult for $D$), we recall a formula for generalized
Laguerre polynomials (see, e.g., \cite{GR-15}) 
\begin{equation}\label{Lna}
L_{n}^{(\alpha)}(x)=
\sum_{k=0}^{n}\binom{n+\alpha}{n-k}\frac{(-x)^k}{k!},
\end{equation}
The special case $L_n(x)\equiv L_n^{(0)}(x)$ is so familiar 
\begin{align}
L_0(x)&=1,
\\
L_1(x)&=-x+1,
\\
L_2(x)&=\frac{x^2}{2}-2x+1,
\\
L_3(x)&=-\frac{x^3}{6}+\frac{3x^2}{2}-3x+1,
\end{align}
that it is easy to recognize them in the basement of half of entries.
The second sequence appearing there, $x+1$, $\frac{1}{2}x^2-1$,
$\frac{1}{6}x^3-\frac{x^2}{2}-x+1$, seem at the first glance not to
fall into the Laguerre case.  However, a simple observation shows that
this also the case:
\begin{align}
L_1(x)-2L_0(x)&=-x-1,
\\
L_2(x)-2L_1(x)&=\frac{x^2}{2}-1,
\\
L_3(x)-2L_2(x)&=-\frac{x^3}{6}+\frac{x^2}{2}+x-1.
\end{align}
To proceed, we recall that the generalized Laguerre polynomials are
subject, among many others, to the relation
\begin{equation}
L_n^{(\alpha-1)}(x)=L_n^{(\alpha)}(x)-L_{n-1}^{(\alpha)}(x).
\end{equation}
We thus meet in our case two linear combinations of the polynomials 
$L_n^{(-1)}(x)$ and $L_{n-1}^{(0)}(x)$, namely, they are 
$L_n^{(-1)}(x)\pm L_{n-1}^{(0)}(x)$.

Our main result here is the following. 

\begin{conjecture}\label{conj}
For $t=1$ and $\Delta=1/2$, and for $r=N-s$, the EFP 
can be given as the determinant of the $s\times s$ matrix 
$I-A$, where $A=DLU$ and    
\begin{equation}\label{DLUij}
\begin{split}
D_{ij}&=h_{r+i}(0)\,\delta_{ij},
\\
L_{ij}&=\frac{(-1)^{i-j}}{h_{r+i}(0)}
\left[L_{i-j}^{(-1)}(\partial_z)+(-1)^{i-1}L_{i-j-1}^{(0)}(\partial_z)\right]
h_{r+i}(z)\Big|_{z=0},
\\ 
U_{ij}&=\frac{(-1)^{i-j}}{h_{r+j}(0)}
\left[L_{j-i}^{(-1)}(\partial_z)+(-1)^{j}L_{j-i-1}^{(0)}(\partial_z)\right]
h_{r+j}(z)\Big|_{z=0}.
\end{split}
\end{equation}
Equivalently, the EFP is given by the determinant of the $s\times s$ matrix 
$I-V$, where $V=\wt L \wt D\wt U$, and 
\begin{equation}\label{wtDLUtij}
\begin{split}
\wt L_{ij}&=\frac{(-1)^{i-j}}{h_{N-i+1}(0)}
\left[L_{i-j}^{(-1)}(\partial_z)+(-1)^{j}L_{i-j-1}^{(0)}(\partial_z)\right]
h_{N-i+1}(z)\Big|_{z=0},
\\ 
\wt D_{ij}&=h_{N-i+1}(0)\,\delta_{ij},
\\
\wt U_{ij}&=\frac{(-1)^{i-j}}{h_{N-j+1}(0)}
\left[L_{j-i}^{(-1)}(\partial_z)+(-1)^{i-1}L_{j-i-1}^{(0)}(\partial_z)\right]
h_{N-j+1}(z)\Big|_{z=0}.
\end{split}
\end{equation}
Function $h_N(z)$ is the Gauss hypergeometric function given
in \eqref{hNz-F1z} or \eqref{hNz-Fz}.
\end{conjecture}
Note that the entries of the matrix $A$ are independent of $s$, if one 
takes $r=N-s$ as an independent geometric parameter.

To verify our conjecture, we have also performed exact evaluation
(with the help of symbolic manipulation software) of the
MIR \eqref{efpMIR1} in the case of $s=5$ for $N\leq 13$. The results
are given in terms of ratios of integers and we find that the
conjectural determinant representation for the EFP reproduces them
exactly.

\section{EFP as a Fredholm determinant}

In this section our aim is to proceed with the determinant formulas
for the EFP and obtain some other representations, specifically, in
the form of Fredholm determinants of linear integral operators.  We
also show that in the scaling limit ($r,s \to \infty$ with their ratio
fixed) some expressions simplify, and, moreover, at the critical value
of $s/r$ corresponding to the arctic curve, there exists local scaling
where the Fredholm determinant turns into the celebrated formula in
terms of the Airy kernel for the Tracy--Widom distribution.

\subsection{Integral form of the matrices}

We start with the explicit formula for the Laguerre polynomials, see
\eqref{Lna}.
It can be easily seen that for a trial function $f(z)$, regular at the
point $z=0$, the following holds:
\begin{equation}\label{LoInt}
\oint_{C_0}\frac{(1-z)^{n+\alpha}}{z^{n+1}}f(z)\frac{\rmd z}{2\pi\rmi}=
(-1)^nL_n^{(\alpha)}(\partial_z)f(z)\big|_{z=0}.
\end{equation}
We are interested in the special cases $\alpha=0$ and $\alpha=-1$.

Reading formula \eqref{LoInt} in the reverse order, for the 
entries of the matrices $L$ and $U$ given in \eqref{DLUij} we can write 
\begin{align}
L_{ij}&=\frac{1}{h_{r+i}(0)}
\oint_{C_0}\left(\frac{(1-z)^{i-j-1}}{z^{i-j+1}}+(-1)^i
\frac{(1-z)^{i-j-1}}{z^{i-j}}\right)
h_{r+i}(z)\frac{\rmd z}{2\pi\rmi},
\\ 
U_{ij}&=\frac{1}{h_{r+j}(0)}
\oint_{C_0}
\left(\frac{(1-z)^{j-i-1}}{z^{j-i+1}}+(-1)^{j+1}
\frac{(1-z)^{j-i-1}}{z^{j-i}}\right)
h_{r+j}(z)\frac{\rmd z}{2\pi\rmi}.
\end{align}

Let us consider now the matrix $A= D L U$, where $D_{ij}=\delta_{ij}
h_{r+j}(0)$, see
\eqref{DLUij}. Slightly simplifying expressions, we get 
\begin{multline}\label{Aijlong}
A_{ij}=\sum_{l=1}^{\min(i,j)} \frac{1}{h_{r+j}(0)}
\oint_{C_0}
\frac{(1-z)^{i-l-1}}{z^{i-l+1}}
\left(1+(-1)^i z\right)h_{r+i}(z)\frac{\rmd z}{2\pi\rmi}
\\ \times
\oint_{C_0}
\frac{(1-w)^{j-l-1}}{w^{j-l+1}}\left(1+(-1)^{j+1} w\right)
h_{r+j}(w)\frac{\rmd w}{2\pi\rmi}.
\end{multline}

Let us now transform expression \eqref{Aijlong}. First, we note that  
the sum over $l$ in \eqref{Aijlong} can be extended to infinity without 
altering the result, since the actual value of the upper limit 
is controlled by the integrals with the respect to $z$ and $w$. 
Next, we evaluate this sum, that yields   
\begin{equation}
\sum_{l=1}^\infty \left(\frac{zw}{(1-z)(1-w)}\right)^l=
\frac{zw}{1-z-w}. 
\end{equation}
Finally, introducing the functions
\begin{equation}
\begin{split}\label{eLU}
e_i^L(z)&=\frac{(1-z)^{i-1}}{z^i}\left(1+(-1)^i z\right)h_{r+i}(z),
\\
e_j^U(w)&=\frac{(1-w)^{j-1}}{h_{r+j}(0)w^j}\left(1+(-1)^{j+1}w\right)h_{r+j}(w),
\end{split}
\end{equation}
for entries of the matrix $A$ we arrive at the formula
\begin{equation}\label{Aijshort}
A_{ij}=\oint_{C_0}\oint_{C_0}
\frac{e_i^L(z)e_j^U(w)}{1-z-w}
\frac{\rmd z\rmd w}{(2\pi\rmi)^2}.
\end{equation}
We will use this formula as the main input for what follows.

As for the matrix $V$, see \eqref{wtDLUtij}, 
one can easily obtain a formula similar to 
\eqref{Aijshort}.

\subsection{Fredholm determinants}

Let $\hat K_\varGamma$ denote a linear integral operator with the
kernel $K(z,w)$ acting on functions on the contour $\varGamma$,
\begin{equation}
\big(\hat K_\varGamma f\big)(z)=\int_\varGamma K(z,w) f(w)\,\rmd w. 
\end{equation} 
The Fredholm determinant of this operator is usually defined as 
\begin{equation}\label{Fredholm}
\det\big(1-\hat K_\varGamma\big)=1+\sum_{n=1}^\infty \frac{(-1)^n}{n!}
\int_\varGamma\dots\int_\varGamma 
\det_{1\leq i,j\leq n}\left[ K(w_i,w_j)\right]\, \rmd w_1\cdots \rmd w_n.
\end{equation}
Another way is to use the identity 
\begin{equation}
\det\big(1-\hat K_\varGamma\big)=
\exp\big\{\tr\log\big(1-\hat K_\varGamma\big)\big\}
\end{equation}
and defining the function
$\log\big(1-\hat K_\varGamma\big)$ by its power series expansion in 
powers of $\hat K_\varGamma$, one can also write
\begin{equation}\label{detexp}
\det\big(1-\hat K_\varGamma\big)=\exp\left\{-\sum_{n=1}^\infty \frac{1}{n} \tr 
\big(\hat K_\varGamma\big)^n
\right\},
\end{equation}
where 
\begin{equation}
\tr \big(\hat K_\varGamma\big)^n=\int_\varGamma\dots \int_\varGamma
K(w_1,w_2)\cdots K(w_n,w_1)\, \rmd w_1\cdots\rmd w_n.
\end{equation}
One can say that the Fredholm determinants of an $s\times s$ matrix
$A$ and linear integral operator $\hat K_\varGamma$ are equal to each
other,
\begin{equation}\label{det=det}
{\det}_s(I-A)=\det\big(1-\hat K_\varGamma\big),
\end{equation}
where the dependence on $s$ is somehow encoded into $\hat
K_\varGamma$, when
\begin{equation}
{\tr}_s A^n=\tr \big(\hat K_\varGamma\big)^n, \qquad n=1,2,\ldots. 
\end{equation}
Here, the subscript $s$ recalls that the matrix $A$ is taken 
to be $s\times s$. 

Let us now consider our matrix $A$ given by \eqref{Aijshort}. We can
immediately claim that we have the identity \eqref{det=det}, where
$\varGamma=C_0$ and the kernel is given by
\begin{equation}\label{Kzz}
K(z_1,z_2)= \sum_{j=1}^s \frac{1}{(2\pi\rmi)^2} 
\oint_{C_0}\frac{e^U_j(w)e_j^L(z_2)}{1-z_1-w}\,\rmd w,\qquad z_1,z_2\in C_0.   
\end{equation}	
Another choice could be with the functions $e_j^U$ and $e_j^L$ being
exchanged.  Note that the dependence on $s$ is encoded into the kernel
in the sum over $j$.

To make our considerations below slightly simpler, we find it useful
at this stage to introduce an integral operator acting on functions on
the real half-axis, $[0,\infty)$. Since $z$ and $w$ lie on the contour
$C_0$, they can be chosen such that $|z|,|w|\ll 1$, and we can use the
formula
\begin{equation}\label{infint}
\frac{1}{1-z-w}=\int_{0}^\infty \rme^{(z+w-1)t}\,\rmd t,\qquad 
\Re (z+w)<1.
\end{equation}
We can  rewrite entries of the matrix $A$ in the form
\begin{equation}
A_{ij}=\int_{0}^{\infty} \rme^{-t} E_i^{L}(t) E_j^{U}(t)\, \rmd t,
\end{equation} 
where we have introduced the functions 
\begin{equation}
E_j^{L,U}(t) =  \oint_{C_0} \rme^{wt} e_j^{L,U}(w)
\,\frac{\rmd w}{2\pi\rmi}.
\end{equation}
As a result, the EFP can be given as the Fredholm determinant 
\begin{equation}\label{FdetKR+}
F_N^{(N-s,s)} =\det\big(1-\hat K_{[0,\infty)}^E\big),  
\end{equation}
where the kernel is 
\begin{equation}
K^E(t_1,t_2) = \rme^{-\frac{1}{2}(t_1+t_2)}\sum_{j=1}^s E_j^L(t_1) E_j^U(t_2).
\end{equation}
More explicitly, 
\begin{equation}\label{KEtt}
K^E(t_1,t_2)= 
\oint_{C_0}\oint_{C_0} \rme^{\left(z-\frac{1}{2}\right)t_1+\left(w-\frac{1}{2}\right)t_2}
\sum_{j=1}^{s} e_j^L(z) e_j^U(w) \frac{\rmd z \rmd w}{(2\pi\rmi)^2}.
\end{equation}
Clearly, the kernel $K^E(t_1,t_2)$ can get a simplified form when the
saddle-point analysis is applied to the contour integrals. This is
what we address below.

\subsection{Scaling limit}

Let us now consider the situation where the size of the system, $N$,
is large.  To have an interesting picture, $r$ and $s$ have to be
taken large as well. We recall that we consider here only the
`symmetric' case, where $r=N-s$.

An important ingredient for subsequent considerations is the following
asymptotic result for the function $h_{N}(z)$.
\begin{proposition}
As $r\to\infty$, 
\begin{equation}\label{hrrC}
h_{r+1}(z)=[\rho(z)]^r C(z)\left(1+O(r^{-1})\right),
\end{equation}
where 
\begin{equation}\label{phoz}
\rho(z)=4\frac{(1-2z)(2-z)(1+z)+2\left(1-z+z^2\right)^{3/2}}{27(1-z)^2}
\end{equation}
and 
\begin{equation}\label{Cz}
C(z)=\frac{2z}{\sqrt{2(1-z+z^2)^2-(2-z)(1-2z)(1+z)\sqrt{1-z+z^2}}}.
\end{equation}
\end{proposition}
We give a proof of this result in Appendix.

Consider now the sum over $j$ in \eqref{KEtt} or in \eqref{Kzz}. From 
\eqref{eLU} we have
\begin{multline}
\sum_{j=1}^s e_j^L(z) e_j^U(w) =
\sum_{j=1}^s [1-z w +(-1)^j (z-w)]
\\ \times
\frac{\left[(1-z)(1-w)\right]^{j-1}}{(zw)^j} 
\frac{h_{r+j}(z)
h_{r+j}(w)}{h_{r+j}(0)}.
\end{multline}
For $r$ large, we can use \eqref{hrrC}. Denoting 
\begin{equation}
\psi(z,w)=\frac{\rho(z)\rho(w)}{\rho(0)},
\end{equation}
we get 
\begin{multline}\label{longexp}
\sum_{j=1}^s e_j^L(z) e_j^U(w) \sim
\frac{C(z)C(w) \left[(1-z)(1-w)\right]^{s-1}[\psi(z,w)]^{r+s-1}}{C(0)(zw)^s} 
\\ \times
\left\{(1-zw)\frac{1-\left(\frac{zw}{(1-z)(1-w)\psi(z,w)}\right)^s}
{1-\frac{zw}{(1-z)(1-w)\psi(z,w)}}
\right. \\ \left. 
+ (-1)^s(w-z)
\frac{1-\left(-\frac{zw}{(1-z)(1-w)\psi(z,w)}\right)^s}
{1+\frac{zw}{(1-z)(1-w)\psi(z,w)}}
\right\}.
\end{multline}
By symbol $\sim$ we denote that the two expressions are equal as
$s,r\to \infty$ up to terms which vanish in the limit; more exactly,
$f\sim g$ means that $f/g\to 1$.

The expression \eqref{longexp} can be simplified 
by noting that $z$ and $w$ are integrated around the 
origin in \eqref{KEtt} and in \eqref{Kzz}. 
Since $\psi(z,w)$ is regular at the points 
$z=0$ and $w=0$, from \eqref{KEtt} we get 
\begin{multline}\label{doubleint}
K^E(t_1,t_2)
\sim
\oint_{C_0}\oint_{C_0}
\rme^{\left(z-\frac{1}{2}\right)t_1+\left(w-\frac{1}{2}\right)t_2}
\\ \times
\frac{C(z)C(w) \left[(1-z)(1-w)\right]^{s-1}[\psi(z,w)]^{r+s-1}}{C(0)(zw)^s} 
\\ \times 
\left\{
\frac{1-zw}
{1-\frac{zw}{(1-z)(1-w)\psi(z,w)}}
+\frac{(-1)^s(w-z)}
{1+\frac{zw}{(1-z)(1-w)\psi(z,w)}}
\right\}
\frac{\rmd z \rmd w}{(2\pi\rmi)^2}.
\end{multline}
When $r$ and $s$ are large, the integrals over $z$ and $w$
can be approximated by the saddle-point 
method. Moreover, both integrals in \eqref{doubleint}
contain as the main factors in their integrands the same function,    
\begin{equation}\label{def-gw}
\left(\frac{1-w}{w}\right)^{s}
\left[\frac{\rho(w)}{\sqrt{\rho(0)}}\right]^N =:
\exp\{Ng(w)\},
\end{equation}
where we have used that $r+s=N$ and included the factor
$\sqrt{\rho(0)}=4/3\sqrt{3}$ for a later convenience (it simplifies a
constant term in $g(w)$).

Let us find the saddle points of the function 
$g(w)$. Denoting 
\begin{equation}
y=\frac{s}{N},
\end{equation}
$y\in(0,1/2]$, we have
\begin{equation}
g(w)=y\log\frac{1-w}{w}+ 
\log \frac{(1-2w)(2-w)(1+w)+2\left(1-w+w^2\right)^{3/2}}{3\sqrt{3}(1-w)^2} .
\end{equation}
We get 
\begin{equation}
g'(w)=\frac{y}{w(w-1)}-\frac{1-\sqrt{1-w+w^2}}{w(w-1)},
\end{equation}
and hence the 
saddle point equation $g'(w)=0$ possesses two solutions $w=w_\pm$, where   
\begin{equation}
w_\pm= \frac{1\pm \sqrt{1-8y+4y^2}}{2}.
\end{equation}
The two solutions collide if $1-8y+4y^2=0$, and choosing the
root of this equation that lies in
the interval $(0,1/2]$, we find that $w_{+}=w_{-}$ when $y=\yc$, where
\begin{equation}\label{yc}
\yc=1-\frac{\sqrt{3}}{2}.
\end{equation}

The critical value $\yc$ separating two asymptotic regimes has a very
clear meaning in terms of the so-called arctic curve. This curve
describes spatial separation between the ordered and disordered phases
of the model in the thermodynamic limit.  In the language of dimer
models, it is the frozen boundary of the limit shape.  For the
six-vertex model with domain wall boundary conditions (with a generic
choice of weights corresponding to $\Delta<1$), it is a curve
inscribed into the unit square, consisting of four portions joining
the four contact points (see \cite{CP-09} for more details).
All the four portions are related to each other by simple symmetry
transformations. They are described in general by a non-algebraic
equation given in a parametric form. For the ice point ($\Delta=1/2$
and $t=1$) this equation appear to be algebraic, and, moreover,
quadratic \cite{CP-08,A-19},
\begin{equation}\label{AC}
4x(1-x)+4y(1-y) +4 xy =1, \qquad x,y\in [0,1/2].
\end{equation}
The arctic curve in the ice-point case is constructed by taking the
arc between the points $(x,y)=(0,1/2)$ and $(x,y)=(1/2,0)$, given
by \eqref{AC}, and applying to it the maps $(x,y)\mapsto (x,1-y)$,
$(x,y)\mapsto (1-x,y)$, and $(x,y)\mapsto (1-x,1-y)$. This yields all
four portions of the arctic curve.

In the context of the EFP, we need only the arc \eqref{AC}.  Recalling
that in the scaling limit the geometric parameters of the EFP scales
as $(N-r)/N=:x$ and $s/N=:y$, we find that the case $r=N-s$ which we
consider here, means simply that $x=y$. Setting $x=y$ in \eqref{AC} we
immediately get the equation $4y^2-8y+1=0$, i.e., we get $y=\yc$ where
$\yc$ is given by \eqref{yc}. Thus the coincidence of the two roots
$w_\pm$ corresponds to the situation where the frozen region of the
EFP touches the arctic curve.  The case $y\in (0,\yc)$ corresponds to
the frozen region of the EFP fully lying in the ordered region,
outside the arctic curve, while the case $y\in (\yc,1/2]$ corresponds
to the frozen region of the EFP partially overlapping with the
disordered region, inside the arctic curve.

The two regimes $y\in (0,\yc)$ and $y\in (\yc, 1/2]$ correspond to two
different behaviours of the EFP.  The values $y\in (0,\yc)$ correspond
to $w_\pm$ real. In this case one finds an exponentially decaying
asymptotic behaviour of the integrals. Among the two points, only the
closest to origin, $w_{-}$, is relevant in the saddle-point
approximation. The values $y\in (\yc, 1/2]$ correspond to $w_\pm$
complex, $w_{-}=w_{+}^*$. In this case both saddle points are relevant
and the leading term in the asymptotic expansion of the integrals is
oscillatory, giving non-vanishing contribution in the limit of large
$N$ (and $s$). Analogous behaviours occur in the context of dimer
models \cite{KOS-06}.

\subsection{Tracy--Widom distribution}

Let us now focus on the vicinity of the point $y=\yc$.  We introduce a
new temporary geometric parameter $\eta$,
\begin{equation}
y=\yc-\eta.
\end{equation}
As $N$ is large, $\eta$ is assumed to be small, below we find that
$\eta \propto N^{-2/3}$.

Let us consider the behaviour of the function $g(w)$ in the vicinity
of the double saddle point $w_{+}=w_{-}=1/2$. 
Setting $w=1/2+\lambda$, and expanding
in the Taylor series in $\lambda$, we find
\begin{equation}
g(w)\Big|_{w=\frac{1}{2}+\lambda}
= 4\eta\lambda-\frac{4}{3\sqrt{3}}\lambda^3+O(\lambda^4).
\end{equation}
Absence of a constant term here is due to the factor $\sqrt{\rho(0)}$
in \eqref{def-gw}.

Hence, setting 
\begin{equation}\label{scaling}
\tilde \lambda=q\lambda, \qquad 
q=\frac{2^{2/3}}{3^{1/6}}N^{1/3},
\end{equation}
and 
\begin{equation}
\sigma= \frac{4N}{q}\eta=2^{4/3}3^{1/6}N^{2/3}\eta,
\end{equation}
we get 
\begin{equation}\label{action}
Ng(w)\Big|_{w=\frac{1}{2}+\lambda}
= \sigma\tilde\lambda-\frac{1}{3}\tilde\lambda^3+O(N^{-1/3}),
\end{equation}
where we write the $O$-term assuming that $\sigma$ and
$\tilde \lambda$ are both of $O(1)$ as $N\to \infty$.  This means that
in the saddle-point approximation we are considering here $\eta\propto
N^{-2/3}$ and $\lambda\propto N^{-1/3}$.  Moreover,
expression \eqref{action} fixes the location of the contour of
integration along the steepest descent directions.

\begin{figure}
\centering
\input{fig-SteepDescent}
\caption{Deformation of the integration contour 
in the $w$-complex plane; shaded area is a vicinity of the 
saddle point at $w=1/2$.}
\label{fig-SteepDescent}
\end{figure}
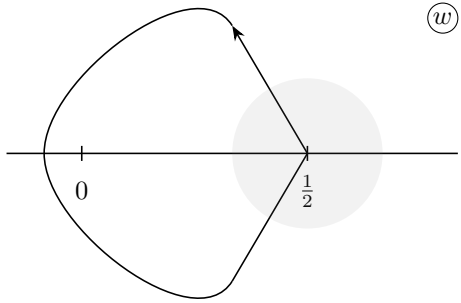

We recall that the initial contour of integration is around the point
$w=0$.  One can deform the contour such that it arrives at the point
$w=1/2$ from direction $-2\pi/3$ and departs from it in the direction
$2\pi/3$, see Fig.~\ref{fig-SteepDescent}. The main contribution to
the integral comes from that part of the deformed contour lying in the
vicinity of the double saddle point $w=1/2$. Let us denote it by
$\gamma$ and let the vicinity of the saddle point be a disk of some
radius $a$ (shown in light grey in the picture). Then,
\begin{equation}
\gamma=(\rme^{-2\rmi\pi/3}a,0)\cup(0,\rme^{2\rmi\pi/3}a). 
\end{equation} 
From \eqref{scaling} it follows that for the scaled variable
$\tilde\lambda$ the contour of integration is $\gamma$, but scaled by
the factor of $q$.  Let us denote this contour by $\tilde\gamma$.
Since $q\to\infty$ as $N\to\infty$, the end-points of $\tilde \gamma$
tend to infinity, and therefore, as it can be shown that up to
exponentially small corrections to the resulting integral,
\begin{equation}\label{tgamma}
\tilde\gamma=(\rme^{-2\rmi\pi/3}\infty,0)\cup(0,\rme^{2\rmi\pi/3}\infty). 	
\end{equation}

Similar considerations hold for the 
integration variable $z$. We 
set $z=1/2+\mu$, and introduce the scaled variable $\tilde \mu$ related to 
$\mu$ by $\mu=q \tilde \mu$, see \eqref{scaling}. 
The contour of integration for $\tilde\mu$ is $\tilde\gamma$.  

Let us now inspect the expression in the braces in \eqref{doubleint}.
Since $\lambda$ and $\mu$ scale as $N^{-1/3}$, we have 
\begin{equation}
\frac{zw}{(1-z)(1-w)\psi(z,w)}
\bigg|_{w=\frac{1}{2}+\lambda, z=\frac{1}{2}+\mu}
=1+2\sqrt{3} (\lambda+\mu)+O(N^{-2/3}).
\end{equation}
The first term in the braces in \eqref{doubleint} behaves as
$1/(\lambda+\mu)=O(N^{1/3})$, and the second term is just of
$O(N^{-1/3})$ thus contributing only to sub-leading terms. As for the
first term, taking also into account the prefactor in the integrand
for which we can use
\begin{equation}
C(1/2)=\frac{2\sqrt{2}}{3},\qquad 
C(0)=\frac{4}{3\sqrt{3}},\qquad \psi(1/2,1/2)=1,
\end{equation}
we get 
\begin{equation}
\frac{C(w)C(z)}{C(0)(1-z)(1-w)\psi(z,w)}
\frac{1-wz}{1-\frac{zw}{(1-z)(1-w)\psi(z,w)}}
\Bigg|_{w=\frac{1}{2}+\lambda, z=\frac{1}{2}+\mu}
=-\frac{1}{\lambda+\mu}+O(1).
\end{equation}
As an intermediate result, we thus have in the leading order  
\begin{equation}
K^E(t_1,t_2)= -\int_{\gamma}\int_{\gamma}
\frac{\rme^{\mu t_1+\lambda t_2 +\sigma (\tilde\lambda+\tilde\mu) 
-(\tilde\lambda^3+\tilde\mu^3)/3} }{\lambda+\mu} 
\frac{\rmd \lambda \rmd \mu}{(2\pi\rmi)^2}.
\end{equation}

To turn to the scaled integration variables $\tilde\lambda$ and
$\tilde\mu$, we change the variables of the kernel, $t_1\mapsto q
t_1$, $t_2\mapsto q t_2$ and introduce the new kernel
\begin{equation}
\wt K^E(t_1,t_2)=q K^E(qt_1,qt_2), \qquad q>0,\qquad  t_1,t_2\in [0,\infty).
\end{equation}
Note that the Fredholm determinants for the kernels $\wt K^E(t_1,t_2)$
and $K^E(t_1,t_2)$ are equal to each other, because this holds for the
traces of arbitrary positive integer powers of the corresponding
integral operators. We have
\begin{equation}
\wt K^E(t_1,t_2)= -\int_{\tilde\gamma}\int_{\tilde\gamma}
\frac{\rme^{\tilde\mu t_1+\tilde\lambda t_2 +\sigma (\tilde\lambda+\tilde\mu) 
-(\tilde\lambda^3+\tilde\mu^3)/3} }{\tilde\lambda+\tilde\mu} 
\frac{\rmd \tilde\lambda \rmd \tilde\mu}{(2\pi\rmi)^2}.
\end{equation}
Summing up, we may write 
\begin{equation}
\lim_{N\to\infty}
\left(F_N^{(N-s,s)}\Big|_{s=N\left(1-\frac{\sqrt{3}}{2}\right)
-\frac{N^{1/3}}{2^{4/3}3^{1/6}}\sigma}\right)=
\det\left(1-\hat{\wt K}{}_{[0,\infty)}^E\right).
\end{equation}
From this point we shall work with these Fredholm determinant and
kernel.

We transform the expression for $\wt K^E(t_1,t_2)$ by noting that the
contour $\tilde\gamma$, see \eqref{tgamma}, can be slightly deformed
such that it passes near the origin ($\tilde \lambda=0$, i.e.,
$w=\frac{1}{2}$) from the left, so that it can be made to lie entirely
in the left half-plane. Note that this is in agreement also with the
condition in \eqref{infint}.  Then, we can use
\begin{equation}
\int_{\sigma}^{\infty} \rme^{(\tilde\lambda+\tilde\mu)t}\,\rmd t= 
- \frac{\rme^{(\tilde\lambda+\tilde\mu)\sigma}}{\tilde\lambda+\tilde\mu},\qquad 
\Re (\tilde\lambda+\tilde\mu) <0, 
\end{equation}
that is valid since $\Re \tilde\lambda<0$ and 
$\Re \tilde\mu<0$ along the contour $\tilde\gamma$. 
We also use the fact that
\begin{equation}
\int_{\tilde\gamma}
\rme^{t \tilde\lambda-\tilde\lambda^3/3}\,
\frac{\rmd \tilde\lambda}{2\pi\rmi}
=\Ai(t),
\end{equation}
where $\Ai(t)$ is the Airy function, that follows from the well-known
contour integral definition of the Airy function in which the change
$\tilde\lambda\mapsto -\tilde\lambda$ has been made.  As a result, for
the kernel $\wt K^E(t_1,t_2)$ we obtain the expression
\begin{equation}
\wt K^E(t_1,t_2)=\int_{\sigma}^{\infty} \Ai(t_1+t)\Ai(t_2+t)\,\rmd t.
\end{equation}

As a final step, we note that 
\begin{equation}
\det\left(1-\hat {\wt K}{}^E_{[0,\infty)}\right)=
\det\left(1-\hat K^{\Ai}_{[\sigma,\infty)}\right)=:\mathcal{F}_2(\sigma),
\end{equation}
where the kernel $K^{\Ai}(t_1,t_2)$ is given by
\begin{equation}\label{Airy-kernel}
K^{\Ai}(t_1,t_2)=\int_{0}^\infty \Ai(t_1+t)\Ai(t_2+t)\,\rmd t,
\end{equation}
and the function $\mathcal{F}_2(\sigma)$ is the celebrated GUE
Tracy--Widom distribution (for the Gaussian unitary ensemble,
$\beta=2$).  The kernel \eqref{Airy-kernel} is also known in the form
\begin{equation}
K^{\Ai}(t_1,t_2)
=\frac{\Ai(t_1)\Ai'(t_2)-\Ai'(t_1)\Ai(t_2)}{t_1-t_2}.
\end{equation}
The proof of equivalence of the two expressions for $K^{\Ai}(t_1,t_2)$
can be found, e.g., in \cite{M-04}.

Summarizing, we conclude that representation $\eqref{DLUij}$ for the
EFP in the six-vertex model with domain wall boundary condition at the
ice point ($\Delta=1/2$ and $t=1$) implies that
\begin{equation}
\lim_{N\to\infty}
\left(F_N^{(N-s,s)}\Big|_{s=N\left(1-\frac{\sqrt{3}}{2}\right)
-\frac{N^{1/3}}{2^{4/3}3^{1/6}}\sigma}\right)=\mathcal{F}_2(\sigma).
\end{equation}
This result is in a full agreement with the one proposed
in \cite{ACJ-23} (see Conjecture 2.5 therein) and with the numerical
simulations in \cite{LKV-23,PS-23}.


\section*{Acknowledgments}

We are indebted to N. M. Bogoliubov, L. Cantini, S. Chhita, H. Spohn,
and J. Viti, for stimulating discussions. AGP is grateful to INFN,
Sezione di Firenze, for hospitality and partial support during his
stay in Florence, Italy, where a part of this work was done.  AGP also
acknowledges partial support from the Theoretical Physics and
Mathematics Advancement Foundation <<BASIS>>.

\appendix
\section{Asymptotic expansion of $h_{r+1}(z)$ as $r\to\infty$}

We have the following expression \cite{CP-08}:
\begin{equation}\label{hrzintegral}
h_{r+1}(z)=\frac{\Gamma(2r+2)}{\left[\Gamma(r+1)\right]^2}
\int_{0}^{1}\left\{t(1-t)\left[1-(1-z)t\right]\right\}^r\,\rmd t.
\end{equation}

Consider first the prefactor in this representation. 
From  the Stirling formula 
\begin{equation}
\log \Gamma(z+a)=\left(z+a-\frac{1}{2}\right)\log z -z
+\log \sqrt{2\pi}+O(1/z)
\end{equation}
we have, as $r\to\infty$, 
\begin{equation}
\frac{\Gamma(2r+2)}{\left[\Gamma(r+1)\right]^2}
= 4^r\sqrt{\frac{4r}{\pi}}\left(1+O(r^{-1})\right).	
\end{equation}

Consider now the integral in \eqref{hrzintegral}. 
The standard saddle-point analysis yields
\begin{equation}
\int_{0}^{1}\left\{t(1-t)\left[1-(1-z)t\right]\right\}^r\,\rmd t=
\rme^{rf(t_0)}\sqrt{\frac{2\pi}{|f''(t_0)|r}}\left(1+O(r^{-1})\right),
\end{equation}
where we have denoted
\begin{equation}
f(t)=\log \big(t (1-t)\left[1-(1-z)t\right]\big),
\end{equation}
and $t_0=t_0(z)$ is the saddle point, $f'(t_0)=0$, given by 
\begin{equation}\label{t0}
t_0=\frac{2-z-\sqrt{1-z+z^2}}{3(1-z)}. 
\end{equation}
Note that $t_0$ lies on the interval of integration
in \eqref{hrzintegral} for broad values of $z$, including the interval
$[0,1]$. We also have
\begin{equation}\label{f''t0}
f''(t_0)=-2\frac{2(1-z+z^2)^2-(2-z)(1-2z)(1+z)\sqrt{1-z+z^2}}{z^2}.
\end{equation}
It is easy to see that $f''(t_0)$ is negative, and so the original
contour goes along the steepest descent directions.  Note that there
exists another solution of the equation $f'(t)=0$, but it is not
relevant (both for analytical and topological reasons).

Collecting things together, we obtain
\begin{equation}
h_{r+1}(z)=[\rho(z)]^r C(z)\left(1+O(r^{-1})\right),
\end{equation}
where
\begin{equation}
\rho(z)=4 \rme^{f(t_0)}= 4t_0(1-t_0)\left[1-(1-z)t_0\right]
\end{equation}
and 
\begin{equation}
C(z)=\sqrt{\frac{8}{|f''(t_0)|}}.
\end{equation}
Substitution of \eqref{t0} yields the expression \eqref{phoz} for the
function $\rho(z)$, and
\eqref{f''t0} leads to \eqref{Cz} for $C(z)$.


\bibliography{freddice_bib}
\end{document}

%% file: fig-SixVertices.tex

\begin{tikzpicture}[scale=.5]
\draw [semithick] (0.2,1)--(1.8,1);
\draw [semithick] (1,0.2)--(1,1.8);
\node at (1,-1) {$a$};
\end{tikzpicture}
\quad
\begin{tikzpicture}[scale=.5]
\draw [line width=2] (0.2,1)--(1.8,1);
\draw [line width=2] (1,0.2)--(1,1.8);
\node at (1,-1) {$a$};
\end{tikzpicture}
\quad
\begin{tikzpicture}[scale=.5]
\draw [semithick] (0.2,1)--(1.8,1);
\draw [line width=2] (1,0.2)--(1,1.8);
\node at (1,-1) {$b$};
\end{tikzpicture}
\quad
\begin{tikzpicture}[scale=.5]
\draw [line width=2] (0.2,1)--(1.8,1);
\draw [semithick] (1,0.2)--(1,1.8);
\node at (1,-1) {$b$};
\end{tikzpicture}
\quad
\begin{tikzpicture}[scale=.5]
\draw [semithick] (0.2,1)--(1,1)--(1,1.8);
\draw [line width=2] (1,0.2)--(1,1);
\draw [line width=2] (1,1)--(1.8,1);
\node at (1,-1) {$c$};
\end{tikzpicture}
\quad
\begin{tikzpicture}[scale=.5]
\draw [line width=2] (0.2,1)--(1,1);
\draw [line width=2] (1,1)--(1,1.8);
\draw [semithick] (1,0.2)--(1,1)--(1.8,1);
\node at (1,-1) {$c$};
\end{tikzpicture}

%% file: fig-DWBCgrid.tex

\begin{tikzpicture}[scale=.5]
\draw [style=dotted, thick] (1,1)--(7,1);
\draw [style=dotted, thick] (1,2)--(7,2);
\draw [style=dotted, thick] (1,3)--(7,3);
\draw [style=dotted, thick] (1,4)--(7,4);
\draw [style=dotted, thick] (1,5)--(7,5);
\draw [style=dotted, thick] (1,6)--(7,6);
\draw [style=dotted, thick] (1,7)--(7,7);
\draw [style=dotted, thick] (1,1)--(1,7);
\draw [style=dotted, thick] (2,1)--(2,7);
\draw [style=dotted, thick] (3,1)--(3,7);
\draw [style=dotted, thick] (4,1)--(4,7);
\draw [style=dotted, thick] (5,1)--(5,7);
\draw [style=dotted, thick] (6,1)--(6,7);
\draw [style=dotted, thick] (7,1)--(7,7);
\draw [semithick] (1,0.5)--(1,1);
\draw [semithick] (2,0.5)--(2,1);
\draw [semithick] (3,0.5)--(3,1);
\draw [semithick] (4,0.5)--(4,1);
\draw [semithick] (5,0.5)--(5,1);
\draw [semithick] (6,0.5)--(6,1);
\draw [semithick] (7,0.5)--(7,1);
\draw [semithick] (7,1)--(7.5,1);
\draw [semithick] (7,2)--(7.5,2);
\draw [semithick] (7,3)--(7.5,3);
\draw [semithick] (7,4)--(7.5,4);
\draw [semithick] (7,5)--(7.5,5);
\draw [semithick] (7,6)--(7.5,6);
\draw [semithick] (7,7)--(7.5,7);
\draw [line width=2] (1,7)--(1,7.5);
\draw [line width=2] (2,7)--(2,7.5);
\draw [line width=2] (3,7)--(3,7.5);
\draw [line width=2] (4,7)--(4,7.5);
\draw [line width=2] (5,7)--(5,7.5);
\draw [line width=2] (6,7)--(6,7.5);
\draw [line width=2] (7,7)--(7,7.5);
\draw [line width=2] (1,1)--(0.5,1);
\draw [line width=2] (1,2)--(0.5,2);
\draw [line width=2] (1,3)--(0.5,3);
\draw [line width=2] (1,4)--(0.5,4);
\draw [line width=2] (1,5)--(0.5,5);
\draw [line width=2] (1,6)--(0.5,6);
\draw [line width=2] (1,7)--(0.5,7);
\end{tikzpicture}

%% file: fig-OnePointBCF.tex



\begin{tikzpicture}[scale=.5]

\draw [style=dotted, thick] (1,1)--(7,1);
\draw [style=dotted, thick] (1,2)--(7,2);
\draw [style=dotted, thick] (1,3)--(7,3);
\draw [style=dotted, thick] (1,4)--(7,4);
\draw [style=dotted, thick] (1,5)--(7,5);
\draw [style=dotted, thick] (1,6)--(7,6);
\draw [style=dotted, thick] (1,1)--(1,6);
\draw [style=dotted, thick] (2,1)--(2,6);
\draw [style=dotted, thick] (3,1)--(3,6);
\draw [style=dotted, thick] (4,1)--(4,6);
\draw [style=dotted, thick] (5,1)--(5,6);
\draw [style=dotted, thick] (6,1)--(6,6);
\draw [style=dotted, thick] (7,1)--(7,6);
\draw [semithick] (1,0.5)--(1,1);
\draw [semithick] (2,0.5)--(2,1);
\draw [semithick] (3,0.5)--(3,1);
\draw [semithick] (4,0.5)--(4,1);
\draw [semithick] (5,0.5)--(5,1);
\draw [semithick] (6,0.5)--(6,1);
\draw [semithick] (7,0.5)--(7,1);
\draw [semithick] (7,1)--(7.5,1);
\draw [semithick] (7,2)--(7.5,2);
\draw [semithick] (7,3)--(7.5,3);
\draw [semithick] (7,4)--(7.5,4);
\draw [semithick] (7,5)--(7.5,5);
\draw [semithick] (7,6)--(7.5,6);
\draw [semithick] (3,7)--(7.5,7);
\draw [line width=2] (1,6)--(1,7.5);
\draw [line width=2] (2,6)--(2,7.5);
\draw [line width=2] (3,7)--(3,7.5);
\draw [line width=2] (4,6)--(4,7.5);
\draw [line width=2] (5,6)--(5,7.5);
\draw [line width=2] (6,6)--(6,7.5);
\draw [line width=2] (7,6)--(7,7.5);
\draw [line width=2] (1,1)--(0.5,1);
\draw [line width=2] (1,2)--(0.5,2);
\draw [line width=2] (1,3)--(0.5,3);
\draw [line width=2] (1,4)--(0.5,4);
\draw [line width=2] (1,5)--(0.5,5);
\draw [line width=2] (1,6)--(0.5,6);
\draw [line width=2] (3,7)--(0.5,7);

\draw [semithick] (3,6)--(3,7);

\draw [decorate,decoration={brace}]
(2.9,8) -- (7.1,8) node [midway,yshift=9pt] {$r$};

\end{tikzpicture}

%% file: fig-EFP.tex


\begin{tikzpicture}[scale=.5]

\draw [style=dotted, thick] (1,1)--(7,1);
\draw [style=dotted, thick] (1,2)--(7,2);
\draw [style=dotted, thick] (1,3)--(7,3);
\draw [style=dotted, thick] (1,4)--(7,4);
\draw [style=dotted, thick] (1,5)--(7,5);
\draw [style=dotted, thick] (1,6)--(7,6);
\draw [style=dotted, thick] (1,7)--(7,7);
\draw [style=dotted, thick] (1,1)--(1,7);
\draw [style=dotted, thick] (2,1)--(2,7);
\draw [style=dotted, thick] (3,1)--(3,7);
\draw [style=dotted, thick] (4,1)--(4,7);
\draw [style=dotted, thick] (5,1)--(5,7);
\draw [style=dotted, thick] (6,1)--(6,7);
\draw [style=dotted, thick] (7,1)--(7,7);
\draw [semithick] (1,0.5)--(1,1);
\draw [semithick] (2,0.5)--(2,1);
\draw [semithick] (3,0.5)--(3,1);
\draw [semithick] (4,0.5)--(4,1);
\draw [semithick] (5,0.5)--(5,1);
\draw [semithick] (6,0.5)--(6,1);
\draw [semithick] (7,0.5)--(7,1);
\draw [semithick] (7,1)--(7.5,1);
\draw [semithick] (7,2)--(7.5,2);
\draw [semithick] (7,3)--(7.5,3);
\draw [semithick] (7,4)--(7.5,4);
\draw [semithick] (7,5)--(7.5,5);
\draw [semithick] (7,6)--(7.5,6);
\draw [semithick] (7,7)--(7.5,7);
\draw [line width=2] (1,5)--(1,7.5);
\draw [line width=2] (2,5)--(2,7.5);
\draw [line width=2] (3,5)--(3,7.5);
\draw [line width=2] (4,7)--(4,7.5);
\draw [line width=2] (5,7)--(5,7.5);
\draw [line width=2] (6,7)--(6,7.5);
\draw [line width=2] (7,7)--(7,7.5);
\draw [line width=2] (1,1)--(0.5,1);
\draw [line width=2] (1,2)--(0.5,2);
\draw [line width=2] (1,3)--(0.5,3);
\draw [line width=2] (1,4)--(0.5,4);
\draw [line width=2] (1,5)--(0.5,5);
\draw [line width=2] (4,6)--(0.5,6);
\draw [line width=2] (4,7)--(0.5,7);

\draw [decorate,decoration={brace}]
(3.9,8) -- (7.1,8) node [midway,yshift=9pt] {$r$};

\draw [decorate,decoration={brace}]
(0,5.9) -- (0,7.1) node [midway,xshift=-9pt] {$s$};

\end{tikzpicture}


%% file: fig-SteepDescent.tex

\usetikzlibrary{arrows.meta}
\begin{tikzpicture}[>=Stealth]
\draw [semithick]  (-1,0)--(5,0);
\draw [semithick] (3,-.1)--(3,.1);
\draw [semithick] (0,-.1)--(0,.1);
\node at (3,-.5) {$\frac{1}{2}$};
\node at (0,-.5) {$0$};
\draw [semithick] [->] (3,0)--(2,1.7);
\draw [semithick] (3,0)--(2,-1.7);
\draw [semithick] (-.5,0) .. controls (-.5,1) and (1.5,2.5) .. (2,1.7);
\draw [semithick] (-.5,0) .. controls (-.5,-1) and (1.5,-2.5) .. (2,-1.7);
\fill [very nearly transparent,gray] (3,0) circle [radius=1];
\node at (4.8,1.8) {$w$};
\draw (4.8,1.8) circle [radius=.2];
\end{tikzpicture}

%% file: freddice_3.bbl
\begin{bibdiv}
\begin{biblist}

\bib{KBI-93}{book}{
      author={Korepin, V.~E.},
      author={Bogoliubov, N.~M.},
      author={Izergin, A.~G.},
       title={Quantum inverse scattering method and correlation functions},
   publisher={Cambridge University Press},
     address={Cambridge},
        date={1993},
}

\bib{BK-02}{incollection}{
      author={Boos, H.~E.},
      author={Korepin, V.~E.},
       title={Evaluation of integrals representing correlations in the {XXX
  H}eisenberg spin chain},
        date={2002},
   booktitle={{MathPhys Odyssey 2001: Integrable Models and Beyond In Honor of
  Barry M. McCoy}},
      editor={Kashiwara, M.},
      editor={Miwa, T.},
      series={Progress in Mathematical Physics},
      volume={23},
   publisher={Birkh{\"a}user Boston, MA},
       pages={65\ndash 108},
      eprint={hep-th/0105144},
         doi={10.1007/978-1-4612-0087-1_4},
}

\bib{BKNS-02}{article}{
      author={Boos, H.~E.},
      author={Korepin, V.~E.},
      author={Nishiyama, Y.},
      author={Shiroishi, M.},
       title={Quantum correlations and number theory},
        date={2002},
     journal={J. Phys. A},
      volume={35},
       pages={4443\ndash 4451},
      eprint={cond-mat/0202346},
         doi={10.1088/0305-4470/35/20/305},
}

\bib{KLNSh-03}{article}{
      author={Korepin, V.E.},
      author={Lukyanov, S.},
      author={Nishiyama, Y.},
      author={Shiroishi, M.},
       title={Asymptotic behavior of the emptiness formation probability in the
  critical phase of {XXZ} spin chain},
        date={2003},
     journal={Phys. Lett. A},
      volume={312},
       pages={21\ndash 26},
      eprint={cond-mat/0210140},
         doi={10.1016/S0375-9601(03)00616-9},
}

\bib{BKS-03}{article}{
      author={Boos, H.~E.},
      author={Korepin, V.~E.},
      author={Smirnov, F.~A.},
       title={Emptiness formation probability and quantum
  {K}nizhnik-{Z}amolodchikov equation},
        date={2003},
     journal={Nucl. Phys. B},
      volume={658},
       pages={417\ndash 439},
      eprint={hep-th/0209246},
         doi={10.1016/S0550-3213(03)00153-6},
}

\bib{GKS-05}{article}{
      author={G\"ohmann, F.},
      author={Kl\"umper, A.},
      author={Seel, A.},
       title={Emptiness formation probability at finite temperature for the
  isotropic {H}eisenberg chain},
        date={2005},
     journal={Physica B},
      volume={359-361},
       pages={807\ndash 809},
      eprint={cond-mat/0406611},
         doi={10.1016/j.physb.2005.01.234},
}

\bib{KKMST-09}{article}{
      author={Kitanine, N.},
      author={Kozlowski, K.~K.},
      author={Maillet, J.~M.},
      author={Slavnov, N.~A.},
      author={Terras, V.},
       title={Algebraic {B}ethe ansatz approach to the asymptotic behavior of
  correlation functions},
        date={2009},
     journal={J. Stat. Mech.: Theor. Exp.},
      volume={2009},
       pages={P04003},
      eprint={0808.0227},
         doi={10.1088/1742-5468/2009/04/P04003},
}

\bib{CP-07b}{article}{
      author={Colomo, F.},
      author={Pronko, A.~G.},
       title={Emptiness formation probability in the domain-wall six-vertex
  model},
        date={2008},
     journal={Nucl. Phys. B},
      volume={798},
       pages={340\ndash 362},
      eprint={0712.1524},
         doi={10.1016/j.nuclphysb.2007.12.016},
}

\bib{CP-08}{article}{
      author={Colomo, F.},
      author={Pronko, A.~G.},
       title={The limit shape of large alternating-sign matrices},
        date={2010},
     journal={SIAM J. Discrete Math.},
      volume={24},
       pages={1558\ndash 1571},
      eprint={0803.2697},
         doi={10.1137/080730639},
}

\bib{CP-09}{article}{
      author={Colomo, F.},
      author={Pronko, A.~G.},
       title={The arctic curve of the domain-wall six-vertex model},
        date={2010},
     journal={J. Stat. Phys.},
      volume={138},
       pages={662\ndash 700},
      eprint={0907.1264},
         doi={10.1007/s10955-009-9902-2},
}

\bib{CS-16}{article}{
      author={Colomo, F.},
      author={Sportiello, A.},
       title={Arctic curves of the six-vertex model on generic domains: the
  tangent method},
        date={2016},
     journal={J. Stat. Phys.},
      volume={164},
       pages={1488\ndash 1523},
      eprint={1605.01388},
         doi={10.1007/s10955-016-1590-0},
}

\bib{CCP-19}{article}{
      author={Cantini, L.},
      author={Colomo, F.},
      author={Pronko, A.~G.},
       title={Integral formulas and antisymmetrization relations for the
  six-vertex model},
        date={2020},
     journal={Ann. Henry Poincar\'e},
      volume={21},
       pages={865\ndash 884},
      eprint={1906.07636},
         doi={10.1007/s00023-019-00856-6},
}

\bib{CGP-21}{article}{
      author={Colomo, F.},
      author={Giulio, G.~Di},
      author={Pronko, A.~G.},
       title={Six-vertex model on a finite lattice: {I}ntegral representations
  for nonlocal correlation functions},
        date={2021},
     journal={Nucl. Phys. B},
      volume={972},
       pages={115535 (42 pp.)},
      eprint={2107.13358},
         doi={10.1016/j.nuclphysb.2021.115535},
}

\bib{TW-94}{article}{
      author={Tracy, C.~A.},
      author={Widom, H.},
       title={Level spacing distributions and the {A}iry kernel},
        date={1994},
     journal={Comm. Math. Phys.},
      volume={159},
       pages={151\ndash 174},
      eprint={hep-th/9211141},
         doi={10.1007/BF02100489},
}

\bib{M-04}{book}{
      author={Mehta, L.~M.},
       title={Random matrices},
     edition={3},
      series={Pure and Applied Mathematics},
   publisher={Elsevier and Academic Press},
     address={Amsterdam},
        date={2004},
      volume={142},
}

\bib{D-06}{incollection}{
      author={Deift, P.},
       title={Universality for mathematical and physical systems},
        date={2007},
   booktitle={{Proceedings of the International Congress of Mathematicians,
  Madrid, 2006}},
      editor={Sanz-Sol\'e, M.},
      editor={Soria, J.},
      editor={Varona, J.~L.},
      editor={Verdera, J.},
   publisher={European Mathematical Society},
     address={Zurich},
       pages={125\ndash 152},
      eprint={math-ph/0603038},
         doi={10.4171/022-1/7},
}

\bib{J-00}{article}{
      author={Johansson, K.},
       title={Shape fluctuations and random matrices},
        date={2000},
     journal={Comm. Math. Phys.},
      volume={209},
       pages={437\ndash 476},
      eprint={math/9903134},
         doi={10.1007/s002200050027},
}

\bib{J-02}{article}{
      author={Johansson, K.},
       title={Non-intersecting paths, random tilings and random matrices},
        date={2002},
     journal={Probab. Theory Relat. Fields},
      volume={123},
       pages={225\ndash 280},
      eprint={math.PR/0011250},
         doi={10.1007/s004400100187},
}

\bib{J-05}{article}{
      author={Johansson, K.},
       title={The arctic circle boundary and the {A}iry process},
        date={2005},
     journal={Ann. Probab.},
      volume={33},
       pages={1\ndash 30},
      eprint={math.PR/0306216},
         doi={10.1214/009117904000000937},
}

\bib{CP-13}{article}{
      author={Colomo, F.},
      author={Pronko, A.~G.},
       title={Third-order phase transition in random tilings},
        date={2013},
     journal={Phys. Rev. E},
      volume={88},
       pages={042125},
      eprint={1306.6207},
         doi={10.1103/PhysRevE.88.042125},
}

\bib{FS-06}{article}{
      author={Ferrari, P.~L.},
      author={Spohn, H.},
       title={Domino tilings and the six-vertex model at its free fermion
  point},
        date={2006},
     journal={J. Phys. A},
      volume={39},
       pages={10297\ndash 10306},
      eprint={cond-mat/0605406},
         doi={10.1088/0305-4470/39/33/003},
}

\bib{KP-16}{article}{
      author={Kitaev, A.~V.},
      author={Pronko, A.~G.},
       title={{Emptiness formation probability of the six-vertex model and the
  sixth Painlev\'e equation}},
        date={2016},
     journal={Comm. Math. Phys.},
      volume={345},
       pages={305\ndash 354},
      eprint={1505.00032},
         doi={10.1007/s00220-016-2636-5},
}

\bib{TW-08a}{article}{
      author={Tracy, C.~A.},
      author={Widom, H.},
       title={Integral formulas for the asymmetric simple exclusion process},
        date={2008},
     journal={Comm. Math. Phys.},
      volume={279},
       pages={815\ndash 844},
      eprint={0704.2633},
         doi={10.1007/s00220-008-0443-3},
}

\bib{TW-08b}{article}{
      author={Tracy, C.~A.},
      author={Widom, H.},
       title={A {F}redholm determinant representation in {ASEP}},
        date={2008},
     journal={J. Stat. Phys.},
      volume={132},
       pages={291\ndash 300},
      eprint={0804.1379},
         doi={10.1007/s10955-008-9562-7},
}

\bib{TW-09}{article}{
      author={Tracy, C.~A.},
      author={Widom, H.},
       title={Asymptotics in {ASEP} with step initial condition},
        date={2009},
     journal={Comm. Math. Phys.},
      volume={290},
       pages={129–154},
      eprint={0807.1713},
         doi={10.1007/s00220-009-0761-0},
}

\bib{BCG-16}{article}{
      author={Borodin, A.},
      author={Corwin, I.},
      author={Gorin, V.},
       title={Stochastic six-vertex model},
        date={2016},
     journal={Duke Math. J.},
      volume={165},
       pages={563\ndash 624},
      eprint={1407.6729},
         doi={10.1215/00127094-3166843},
}

\bib{LKV-23}{article}{
      author={Lyberg, I.},
      author={Korepin, V.},
      author={Viti, J.},
       title={Fluctuation of the phase boundary in the six-vertex model with
  domain wall boundary conditions: a {Monte Carlo} study},
        date={2023},
     journal={J. Phys. A: Math. Theor.},
      volume={56},
       pages={495002},
      eprint={2303.14669},
         doi={10.1088/1751-8121/ad0a43},
}

\bib{PS-23}{article}{
      author={Pr\"ahofer, M.},
      author={Spohn, H.},
       title={Domain wall fluctuations of the six-vertex model at the ice
  point},
        date={2023},
     journal={J. Phys. A: Math. Theor.},
      volume={57},
       pages={025001},
      eprint={2305.09502},
         doi={10.1088/1751-8121/ad13b4},
}

\bib{ACJ-23}{article}{
      author={Ayyer, A.},
      author={Chhita, S.},
      author={Johansson, K.},
       title={{GOE} fluctuations for the maximum of the top path in alternating
  sign matrices},
        date={2023},
     journal={Duke Math. J.},
      volume={172},
       pages={1961\ndash 2104},
      eprint={2109.02422},
         doi={10.1215/00127094-2022-0075},
}

\bib{B-82}{book}{
      author={Baxter, R.~J.},
       title={Exactly solved models in statistical mechanics},
   publisher={Academic Press},
     address={San Diego, CA},
        date={1982},
}

\bib{K-82}{article}{
      author={Korepin, V.~E.},
       title={Calculations of norms of {B}ethe wave functions},
        date={1982},
     journal={Commun. Math. Phys.},
      volume={86},
       pages={391\ndash 418},
         doi={10.1007/BF01212176},
}

\bib{I-87}{article}{
      author={Izergin, A.~G.},
       title={Partition function of the six-vertex model in a finite volume},
        date={1987},
     journal={Sov. Phys. Dokl.},
      volume={32},
       pages={878\ndash 879},
}

\bib{ICK-92}{article}{
      author={Izergin, A.~G.},
      author={Coker, D.~A.},
      author={Korepin, V.~E.},
       title={Determinant formula for the six-vertex model},
        date={1992},
     journal={J. Phys. A: Math. Gen.},
      volume={25},
       pages={4315\ndash 4334},
         doi={10.1088/0305-4470/25/16/010},
}

\bib{EKLP-92a}{article}{
      author={Elkies, N.},
      author={Kuperberg, G.},
      author={Larsen, M.},
      author={Propp, J.},
       title={Alternating-sign matrices and domino tilings ({Part 1})},
        date={1992},
     journal={J. Algebraic Combin.},
      volume={1},
       pages={111\ndash 132},
      eprint={math/9201305},
         doi={10.1023/A:1022420103267},
}

\bib{EKLP-92b}{article}{
      author={Elkies, N.},
      author={Kuperberg, G.},
      author={Larsen, M.},
      author={Propp, J.},
       title={Alternating-sign matrices and domino tilings ({Part 2})},
        date={1992},
     journal={J. Algebraic Combin.},
      volume={1},
       pages={219\ndash 234},
      eprint={math/9201305},
         doi={10.1023/A:1022483817303},
}

\bib{Ku-96}{article}{
      author={Kuperberg, G.},
       title={Another proof of the alternative-sign matrix conjecture},
        date={1996},
     journal={Int. Math. Res. Not.},
      volume={1996},
       pages={139\ndash 150},
      eprint={math/9712207},
         doi={10.1155/S1073792896000128},
}

\bib{Ze-96}{article}{
      author={Zeilberger, D.},
       title={Proof of the refined alternating sign matrix conjecture},
        date={1996},
     journal={New York J. Math.},
      volume={2},
       pages={59\ndash 68},
      eprint={math/9606224},
}

\bib{Br-99}{book}{
      author={Bressoud, D.~M.},
       title={Proofs and confirmations: The story of the alternating sign
  matrix conjecture},
   publisher={Cambridge University Press},
     address={Cambridge},
        date={1999},
}

\bib{BPZ-02}{article}{
      author={Bogoliubov, N.~M.},
      author={Pronko, A.~G.},
      author={Zvonarev, M.~B.},
       title={Boundary correlation functions of the six-vertex model},
        date={2002},
     journal={J. Phys. A: Math. Gen.},
      volume={35},
       pages={5525\ndash 5541},
      eprint={math-ph/0203025},
         doi={10.1088/0305-4470/35/27/301},
}

\bib{S-06}{article}{
      author={Stroganov, Yu.~G.},
       title={Izergin--{K}orepin determinant at a third root of unity},
        date={2006},
     journal={Theor. Math. Phys.},
      volume={146},
       pages={53\ndash 62},
      eprint={math-ph/0204042},
         doi={10.1007/s11232-006-0006-8},
}

\bib{PP-19}{article}{
      author={Pronko, A.~G.},
      author={Pronko, G.~P.},
       title={Off-shell {B}ethe states and the six-vertex model},
        date={2019},
     journal={J. Math. Sci.},
      volume={242},
       pages={742\ndash 752},
         doi={10.1007/s10958-019-04511-7},
}

\bib{CP-05a}{article}{
      author={Colomo, F.},
      author={Pronko, A.~G.},
       title={On the refined $3$-enumeration of alternating sign matrices},
        date={2005},
     journal={Adv. in Appl. Math.},
      volume={34},
       pages={798\ndash 811},
      eprint={math-ph/0404045},
         doi={10.1016/j.aam.2004.09.007},
}

\bib{STW-22}{incollection}{
      author={Saenz, A.},
      author={Tracy, C.~A.},
      author={Widom, H.},
       title={Domain walls in the {H}eisenberg--{I}sing spin-$\frac{1}{2}$
  chain},
        date={2022},
   booktitle={{Toeplitz Operators and Random Matrices: In Memory of Harold
  Widom}},
      editor={Basor, E.},
      editor={B{\"o}ttcher, A.},
      editor={Ehrhardt, T.},
      editor={Tracy, C.~A.},
      series={Operator Theory: Advances and Applications},
      volume={289},
   publisher={Birkh{\"a}user, Cham},
       pages={9\ndash 47},
      eprint={2202.07695},
         doi={10.1007/978-3-031-13851-5_2},
}

\bib{GR-15}{book}{
      author={Gradshteyn, I.~S.},
      author={Ryzhik, I.~M.},
       title={Table of integrals, series, and products},
     edition={8th ed.},
   publisher={Academic Press},
        date={2015},
}

\bib{A-19}{article}{
      author={Aggarwal, A.},
       title={Arctic boundaries of the ice model on three-bundle domains},
        date={2019},
     journal={Invent. Math.},
      volume={220},
       pages={611\ndash 671},
      eprint={1812.03847},
         doi={10.1007/s00222-019-00938-6},
}

\bib{KOS-06}{article}{
      author={Kenyon, R.},
      author={Okounkov, A.},
      author={Sheffield, S.},
       title={Dimers and amoebae},
        date={2006},
     journal={Ann. of Math.},
      volume={163},
       pages={1019\ndash 1056},
      eprint={math-ph/0311005},
         doi={10.4007/annals.2006.163.1019},
}

\end{biblist}
\end{bibdiv}
